\documentclass[npg]{copernicus}
\begin{document}
\title{Distinguishing the effects of internal and forced atmospheric variability in climate networks}
\author[1]{J.~I. ~Deza}
\author[2]{C. ~Masoller}
\author[3]{M. ~Barreiro}
\affil[1]{Departament de F\'isica i Enginyeria Nuclear, Universitat Polit\`ecnica de Catalunya, Colom 11. E-08222, Terrassa, Barcelona, Spain.}
\affil[2]{Instituto de F\'isica, Facultad de Ciencias, Universidad de la Rep\'ublica, Igu\`a 4225, Montevideo, Uruguay.}
 \correspondence{juan.ignacio.deza@upc.edu.}
\date{\today}
\maketitle

\begin{abstract}
 The fact that the Earth climate is a highly complex dynamical system is well-known. In the last few decades a lot of effort has been focused on understanding how climate phenomena in one geographical region affects the climate of other regions. Complex networks are a powerful framework for identifying climate interdependencies. To further exploit the knowledge of the links uncovered via the network analysis (for, e.g., improvements in prediction), a good understanding of the physical mechanisms underlying these links is required. Here we focus in understanding the role of atmospheric variability, and construct climate networks representing internal and forced variability. In the connectivity of these networks we assess the influence of two main indices, NINO3.4 and the North Atlantic Oscillation (NAO), by calculating the networks from time-series where these indices were linearly removed. We find that the connectivity of the forced variability network is heavily affected by ``El Ni\~no'': removing the NINO3.4 index yields a general loss of connectivity; even teleconnections between regions far away from the equatorial Pacific ocean are lost, suggesting that these regions are not directly linked, but rather, are indirectly interconnected via ``El Ni\~no'', particularly on interannual time scales. On the contrary, in the internal variability network (independent of sea surface temperature forcing) we find that the links are significantly affected by NAO with a maximum in intra-annual time scales. While the strongest non-local links found are those forced by the ocean, we show that there are also strong teleconnections due to internal atmospheric variability.
\end{abstract}

\introduction
The existence of long-range teleconnections in the Earth climate is well-known, as the atmosphere connects far away regions through waves and advection of heat and momentum. This long-range coupling makes the complex network approach \citep{albert_barabasi_rmp} of the Earth's climate very attractive \citep{tsonisWhat2006,tsonis_telcon_2008,DongesBackbone2009,zou2011recent}. Climate networks are constructed by considering the Earth as a regular grid of nodes and assigning links connecting two different nodes via an analysis of their similarity over a particular field. This approach has been used in the literature both on local and on global scales and for analyzing climate phenomena considering both linear and nonlinear interdependencies.

For example, the network approach has been recently used to analyze patterns of extreme monsoonal rainfall over South Asia \citep{malik2012Analysis}, to infer early warning indicators for the Atlantic Meridional Overturning Circulation collapse \citep{early_warning_indicators}, to gain insight into the  origin of decadal climate variability \citep{Tsonis_decadal_variability} and to study El Ni{\~n}o phenomenon as an autonomous component in the climate system \citep{gozolchiani2011emergence}.

Various methods for constructing climate networks have been proposed (computing information measures from temperature or geopotential fields, from daily or monthly data, etc.) and the reliability and robustness of the networks uncovered have also been analyzed in terms of a critical comparison of the networks found with the various methods used \citep{palusDiscerning,palusReliability,Davidsen_interpretation,giulio}.

Here we focus in understanding the role of atmospheric variability in the climate by means of networks constructed from monthly averaged surface air temperature (SAT) anomalies.

Atmospheric variability can be considered, to first order, as a superposition of an internal part due to intrinsic local dynamics, and an external part due to the variations of the boundary conditions, primarily given by the sea surface temperature (SST) forcing. These two components can be distinguished by using Atmospheric General Circulation Models (AGCMs) forced with prescribed historical SSTs \citep{straus_2000,barreiro_2002,Molteni2003Atmospheric, KucharskiInternal} (see also the accompanying paper in this Special Issue).

Separating forced from internal atmospheric variability is important because it can allow for improvements in climate prediction. In many geographical regions the climate is strongly influence by SST variations that force persistent anomalies in the atmosphere \citep{ShuklaPredictability}. Because the evolution of the tropical oceans is itself predictable, this allows prediction of atmospheric variables beyond the chaotic time scale of 7-10 days due to the sensitivity to initial conditions \citep{nature_2012}.

The usual modeling strategy to study predictability consists on forcing AGCMs with idealized or observed SST anomalies. This allows investigating the response of the atmosphere to different boundary conditions and different initial conditions. If we consider the time series of anomalies of a climatic field (let's say, SAT anomalies) as a combination of internal and forced variability, e.g. $x=x_{for}+x_{int}$, we can use the output of several numerical experiments initialized differently but forced with the \emph {same} boundary conditions (i.e. same SST) to separate the internal and forced variability. For each run $i$ we have $$x^i=x^i_{for}+x^i_{int} = x_{for}+x^i_{int}$$ (as $x_{for}$ does not depend on the initial conditions).
Averaging over $N$ runs yields $$\bar{x}= x_{for}+ (1/N)\sum_i x^i_{int}.$$ If the average is done over several model runs, the second term is washed out and $\bar{x}\approx x_{for}$.

In other words, each time series $x^i$ can be broken down into a part that changes from run to run, $x^i_{int}$, and a part that does not depend on the initial conditions (is forced by the boundary conditions only and is the same for all runs), $x_{for}\approx \bar{x}$.

We use this method to construct two types of networks, those in which the links represent similarities in internal atmospheric variability (referred as {\it internal variability} network), and those in which the links represent similarities in forced atmospheric variability (the {\it forced variability} network).

In the connectivity of these networks we assess the influence of two main phenomena: El Ni{\~n}o characterized by the NINO3.4 index and the North Atlantic Oscillation, characterized by the NAO index. We do so calculating the networks from time-series where either the NINO3.4 index or the NAO index was linearly removed.

We find that the forced variability networks are intimately related to ``El Ni\~no'' phenomenon and that linearly removing its evolution yields a breakdown of the long range teleconnections of the climate network, particularly in interannual time scales. A similar result is observed for the internal variability network in the Northern Hemisphere when NAO is removed, with maximum effect in intra-annual time scales.

The paper is organized as follows. In Section II the NINO3.4 and NAO indices are described. In Section III the method used for constructing climate networks is shown. The data and the model employed are discussed in Section IV. Section V presents the results. The internal and forced variability networks, and the effects of NAO and ``El Ni\~no'' are analyzed. Section VI presents a summary and the conclusions.

\section{Climate indices}
A climate index describes the state and changes of a particular region of the ocean or the atmosphere. Indices can be determined from monitoring station data or identified by means of Empirical Orthogonal Functions (EOF) analysis. In the latter case, they result as the principal component (PC) related to the strongest EOF over a chosen area, calculated for a pre-determined variable (e.g. temperature or pressure).  Here we use the second approach to compute the indices.

\subsection{NINO 3.4}
The NINO 3.4 index \citep{TrenberthDefinition} is calculated as the average of SST anomalies in the equatorial Pacific bounded by latitudes $5S-5N$ and by longitudes $120W-170W$. The monthly index from NOAA \citep{indicesData}, updated monthly has been used from 1948 to 2006. As this index is based on SST, a boundary condition on the AGCM, it is clear that this phenomenon will affect mainly the \emph{forced} part of the atmospheric variability. 
\subsection{NAO}

The North Atlantic Oscillation (NAO) has been shown to be mainly an \emph{atmospheric} phenomenon not forced by the ocean\citep{HurrellDecadal}.
The NAO index is calculated as the leading EOF of surface pressure over the north Atlantic region ($20N-80N$ and  90W-40E) for each model run. Comparison among indices from different model runs and between these and the observed NAO index from reanalysis data yielded different time series modulating essentially the same spatial pattern. These series have the same statistical properties of low frequency noise.

\section{Methods for network construction}

\subsection{Measure of statistical interdependence}
The Mutual information (MI) is computed from the probability density functions (PDFs) that characterize two time series in two nodes, $p_i$ and $p_j$, as well as their joint probability function, $p_{ij}$\citep{Palus2007ContempPhysics}:
\begin{equation}
M_{ij}=\sum_{m,n}  p_{ij}(m,n) \log \frac{p_{ij}(m,n)}{p_i(m) p_j(n)}.
\end{equation}
$M_{ij}$ is a symmetric measure ($M_{ij}=M_{ji}$) of the degree of statistical interdependence of the time series in nodes $i$ and $j$; if they are independent:  $p_{ij}(m,n) = p_i(m) p_j(n)$ and $M_{ij}=0$.

In this paper the PDFs $p_i$, $p_j$ and $p_{ij}$ are computed in two ways: by usual histograms of values (in the following, when the probabilities are estimated with histograms of values, the MI will be referred to as MIH) and by using a symbolic transformation, in terms of probabilities of {\it ordinal patterns} (in the following, the MI computed from probabilities of patterns will be referred to as MI OP)\citep{Bandt2002Permutation,barreiro2011inferring}.

The ordinal patterns are calculated from time series by comparing the value of a given data point relative to its neighbors. When a value ($v_2$) is higher than the previous one ($v_1$) and lower than the next one ($v_3$) ($v_1 < v_2 < v_3$), the ordinal transformation gives pattern ``123''; when $v_1 > v_2 > v_3$, it gives pattern ``321'', and so on. Considering patterns of length $D$, then there are $D!$ possible patterns.

This symbolic transformation keeps the information about correlations present in a time-series, but does not keep information about the absolute values of the data points. Therefore, the symbolic mutual information (MI OP) can be expected to provide complementary information with respect to the usual way of computing the mutual information (MIH).

A significant advantage for climate data analysis is that the ordinal transformation allows for tuning the time-scale of the analysis, not only by considering shorter or longer patterns, but also, by comparing data points in the time-series which are not consecutive but separated by a lag.

As we analyze monthly data in the period January 1948 - December 2006, due to the short length of the time series ($708$ data points), in order to compute the probabilities of the patterns with good statistics we considered ordinal patterns of length 3. Since there are 6 possible patterns of length 3, for the sake of consistency, the MIH is computed using 6-bin histograms.

We varied the time-scale of the MI OP analysis by constructing the patterns in three ways: 1) by comparing temperature anomalies in 3 consecutive months (constructing patterns with 3 consecutive data points), 2) by comparing anomalies in 3 consecutive seasons (by taking one data point every 4 points) and 3) by comparing anomalies in the same month of 3 consecutive years (by taking one data point every 12 points). The MI OP computed in these ways is referred to as {\it intra-seasonal}, {\it intra-annual} and  {\it inter-annual} respectively.

\subsection{Thresholding}

To construct the network we consider that there is a link between nodes $i$ and $j$ if $M_{ij}$ is above an appropriate threshold, which is calculated in terms of {\it surrogated} shuffled data \citep{deza2013Inferring}. We calculate the mean, $\mu$, and the standard deviation, $\sigma$, of the distribution of MI values computed from surrogated data, and accept links whose MI value, computed from the original data, is above $\mu + 3 \sigma$.

\subsection{Graphical representation}

To represent the network we plot the usual {\em area-weighted connectivity} (AWC) of the nodes, which is the fraction of the total area of Earth to which each node is connected, that is
\begin{equation}
AWC_{i}=\frac{\sum_{j}^N A_{ij} \cos(\lambda_{j})}{\sum_{j}^N \cos(\lambda_{j})},
\label{awc}
\end{equation}
where $\lambda_{i}$ is the latitude of node $i$ and $A_{ij}=1$ if nodes $i$ and $j$ are connected and zero otherwise. 
As we are particularly interested in identifying significant weak links, in all the AWC maps presented in Sec. V, the color scale has been set from zero to a fix value ($0.4$), and any node with stronger connectivity is shown with the color code of 0.4.

In Sec. V we also analyze the connections of a few selected geographical regions (represented by individual network nodes) and in these connectivity maps we display the value of the interdependency measure (MIH or MI OP), using a color scale that is also fixed, from zero up to $0.3$; MI values larger than this are shown with the same color code as 0.3.

To summarize, the graphical network representation is done by plotting AWC maps and node connectivity maps, and for easier comparison, the color code is kept fixed (0-0.4 in the AWC maps; 0-0.3 in the connectivity maps).

\begin{figure*}[]
\begin{center}
\includegraphics[width=0.49\textwidth]{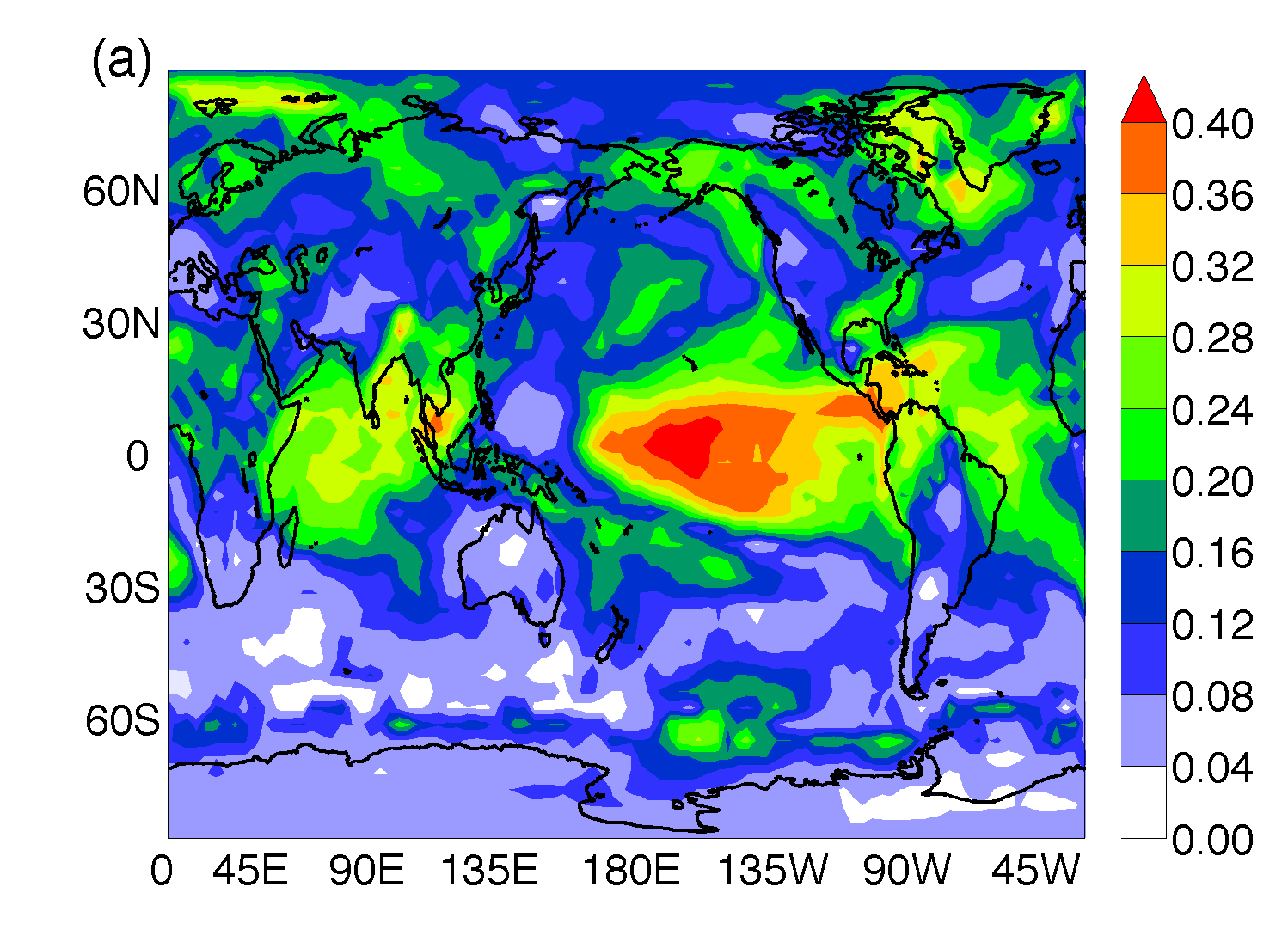}
\includegraphics[width=0.49\textwidth]{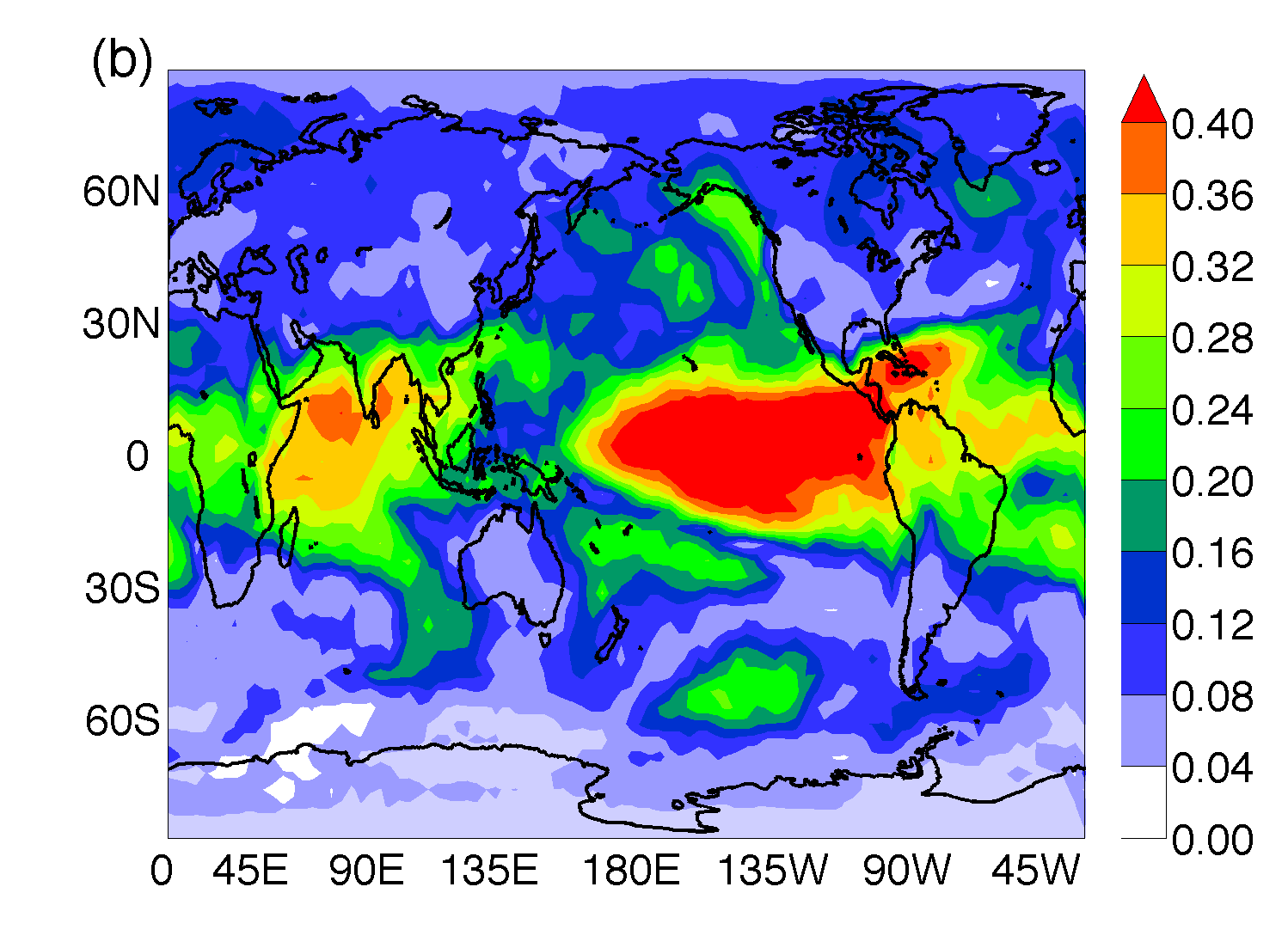}
\caption{
Maps of AWC (see Eq. \ref{awc}) constructed from reanalysis NCEP/NCAR data. The statistical interdependencies are quantified via (a) MIH and (b) MI OP inter-annual time-scale (see Sec. III A for details). The color scale is the same for both panels and for all the following AWC maps.} \label{fig:reanalysis}
\end{center}
\end{figure*}

\section{Data sets and model used}
In this study we used the AGCM from the International Centre for Theoretical Physics (ICTP AGCM), a full atmospheric model with simplified physics and an horizontal resolution of T30 ($3.75^\circ \times 3.75^\circ$, which gives $N = 608$ grid points or network nodes) with eight vertical levels\citep{Molteni2003Atmospheric}. The model is forced with historical global sea surface temperatures (ERSSTv.2)\citep{smith2004Improved}. In order to separate forced from internal atmospheric variability nine runs using the same boundary (SSTs) conditions but slightly different initial conditions were performed.

In our experiment design SST is taken as a boundary condition and it is not changed by the atmospheric flow. In the real world there is a two-way interaction between the ocean and the atmosphere. This limitation is especially important in the extra-tropics where the SST evolution strongly depends on the atmospheric forcing. However, current understanding indicates that the atmosphere is most sensitive to SST anomalies in the tropics and thus the forced atmospheric variability will be related to the evolution of the tropical oceans. Thus this model setup allows, as explained in the introduction, to separate the \emph{ forced} and \emph {internal} components of the atmospheric variability. As we are using only nine runs from the model, a perfect separation is not to be expected, however this number of runs has been shown enough to study forced variability in many regions of the world\citep{rodwell1999Oceanic,barreiro2009Influence}.

We analyze monthly averaged air surface temperature in the period January 1948 - December 2006, and thus, in each node we have a time series of $708$ data points. In each node we computed the anomalies (by subtracting the average for each month) and the time series was linearly detrended and normalized by the standard deviation.

For assessing the importance of the NINO3.4 or the NAO indices, we constructed networks from time series from which we have previously removed the evolution of NINO3.4 or NAO. This was done in two steps: 1) linearly regress the time series of each node with respect to the time series of the index; 2) subtract the linear regression from the original data. This procedure removes effectively the linear contribution of the index considered in the evolution of each node.

To validate the model (see Section V A) we considered reanalisis data from NCEP/NCAR \citep{Kalnay1996NMC/NCAR} in the same time period (1948-2006). Since NCEP/NCAR reanalysis data is given on a $2.5^\circ \times 2.5^\circ$ grid, for easier comparison it was resampled to fit the grid of the ICTP-AGCM data.

\section{Results}
\begin{figure*}[]
\begin{center}
\includegraphics[width=0.49\textwidth]{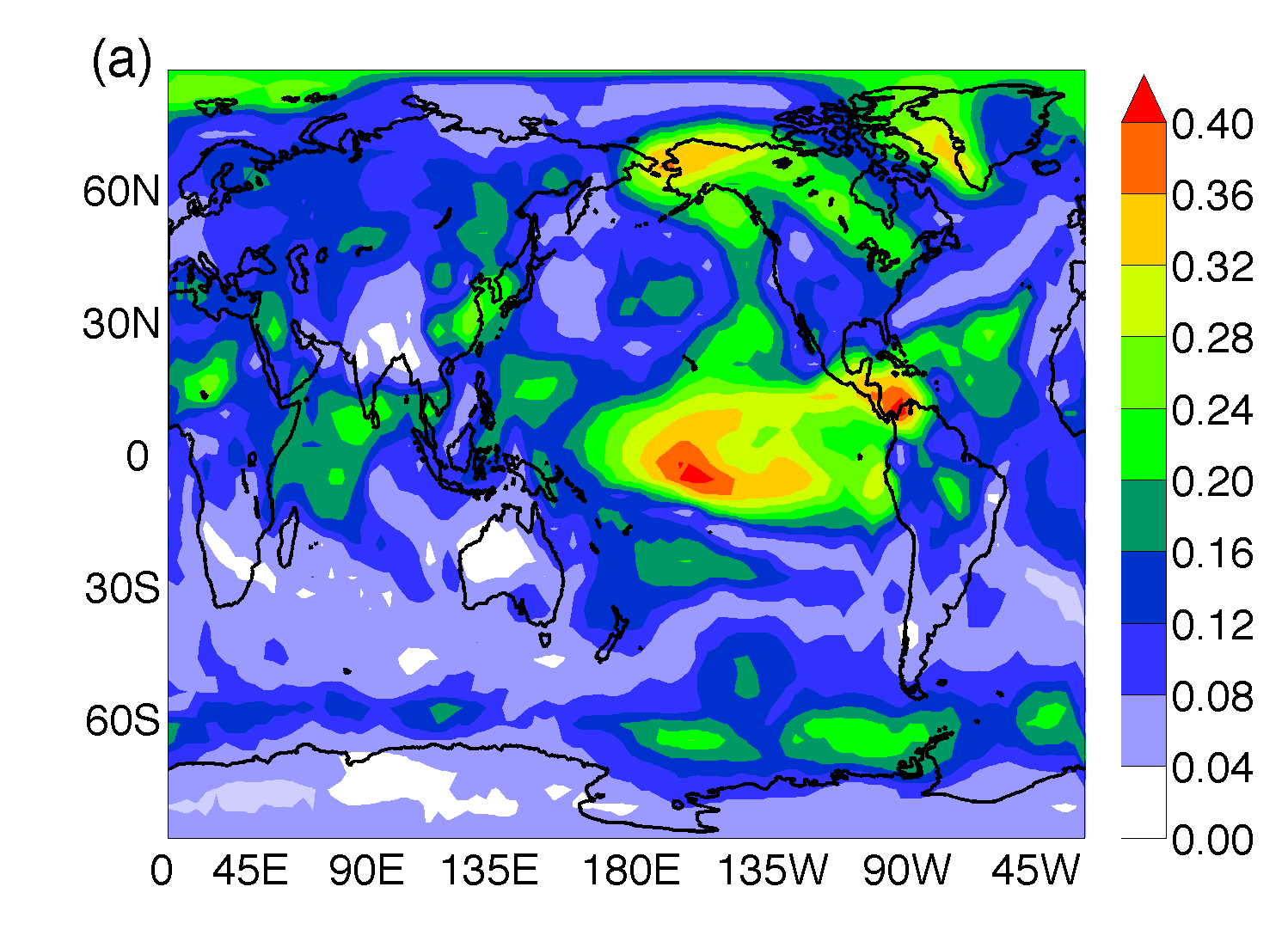}
\includegraphics[width=0.49\textwidth]{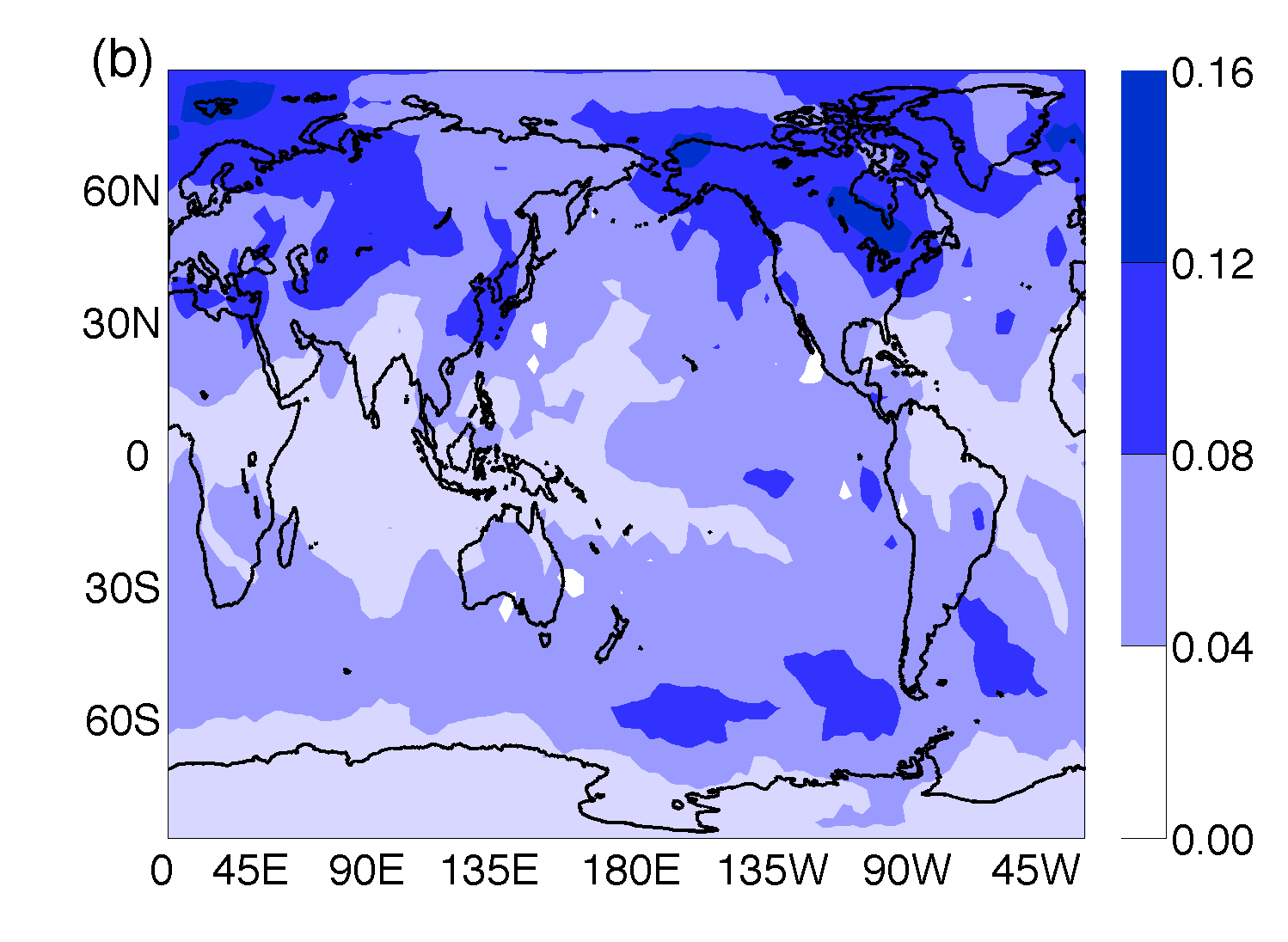}
\includegraphics[width=0.49\textwidth]{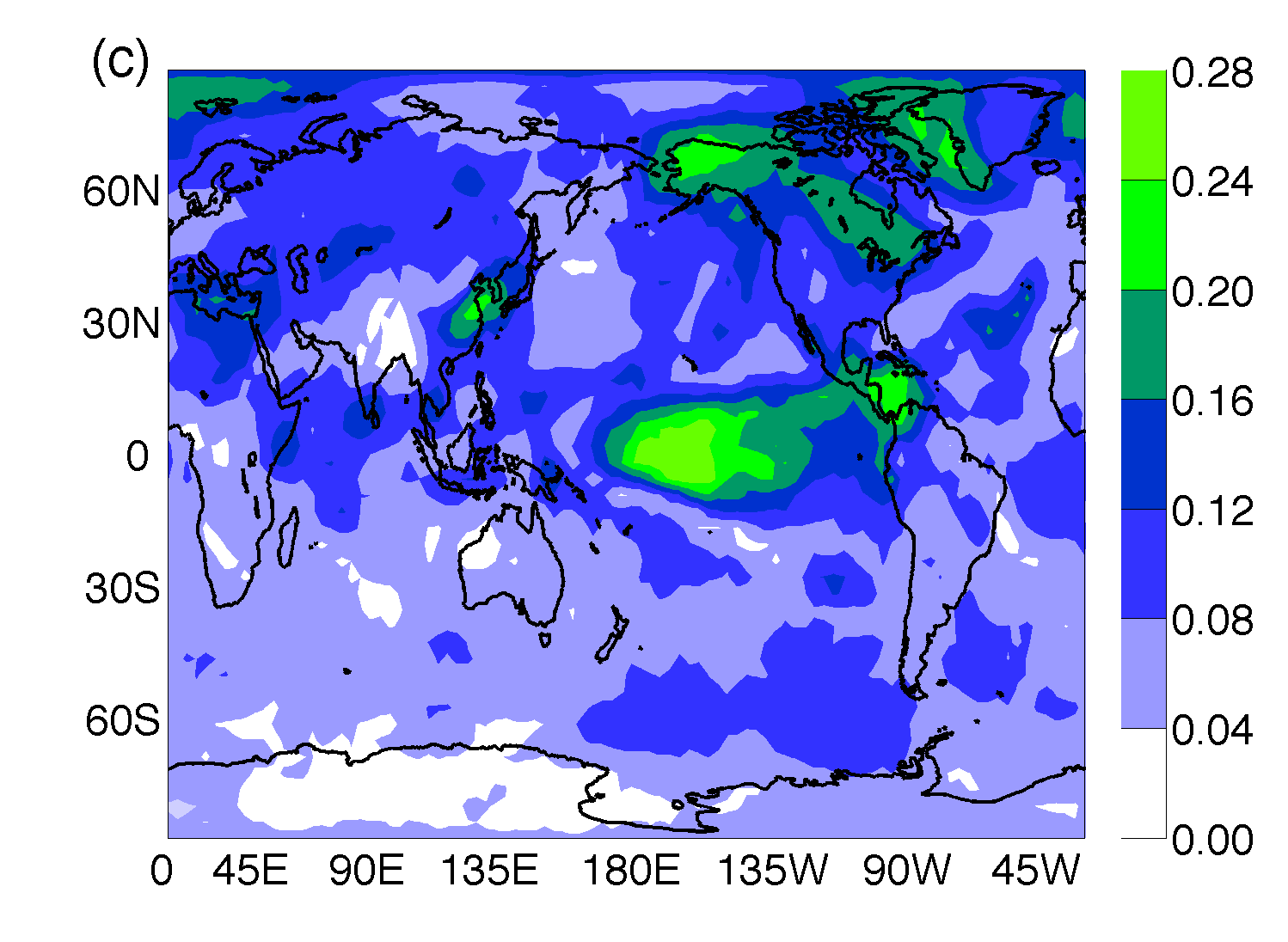}
\includegraphics[width=0.49\textwidth]{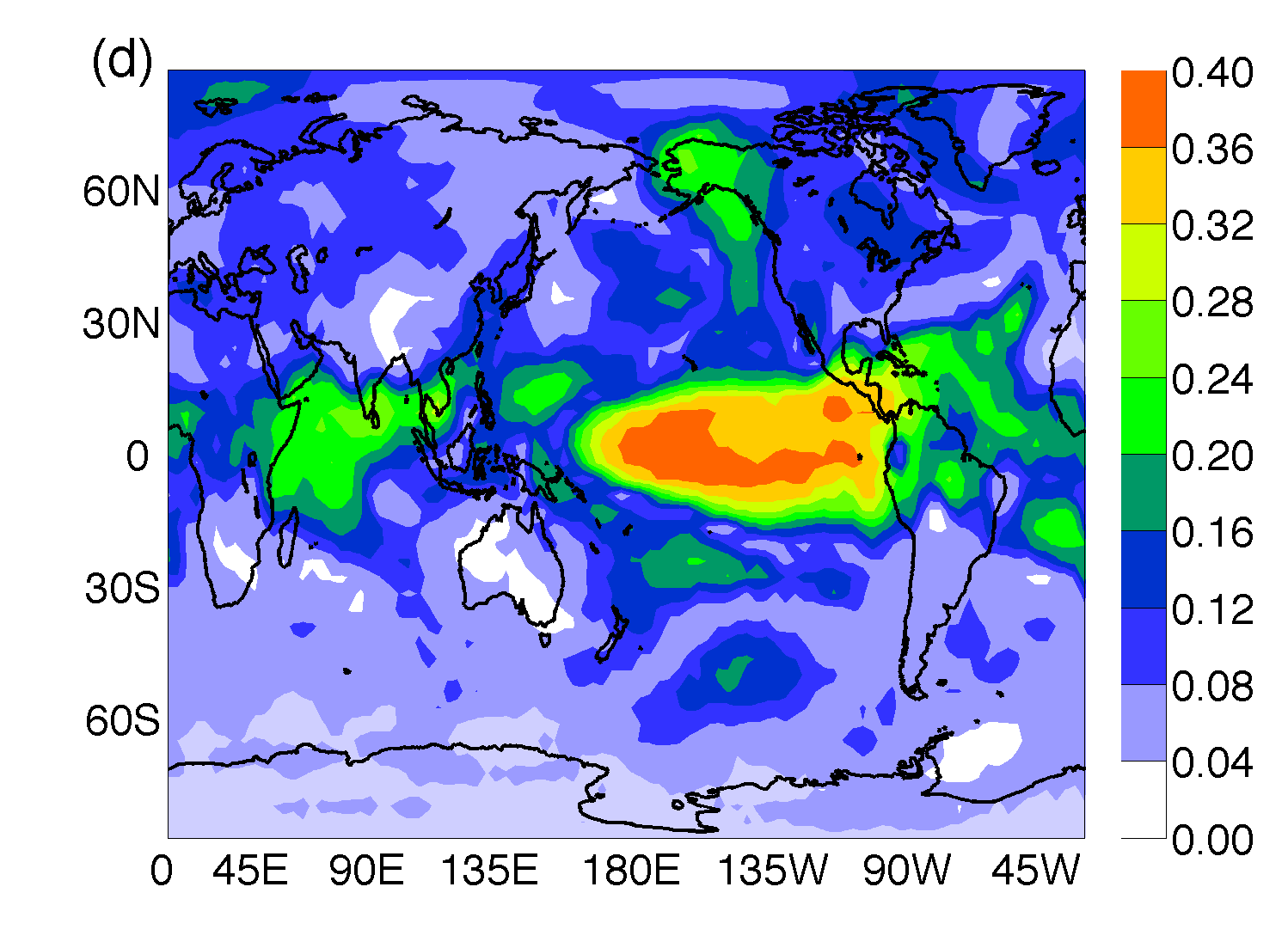}
\caption{Maps of AWC obtained from single model run. The statistical interdependencies are quantified via (a) MIH, (b) MI OP intra-season, (c) intra-annual and (d) inter-annual  (see Sec. III A for details). Comparing panel (a) with Fig. \ref{fig:reanalysis}(a) and panel (d) with Fig. \ref{fig:reanalysis}(b) we observe that the main features of the maps are the same, providing a visual validation of the model. } \label{fig:full}
\end{center}
\end{figure*}

\begin{figure*}[]
\begin{center}
\includegraphics[width=0.49\textwidth]{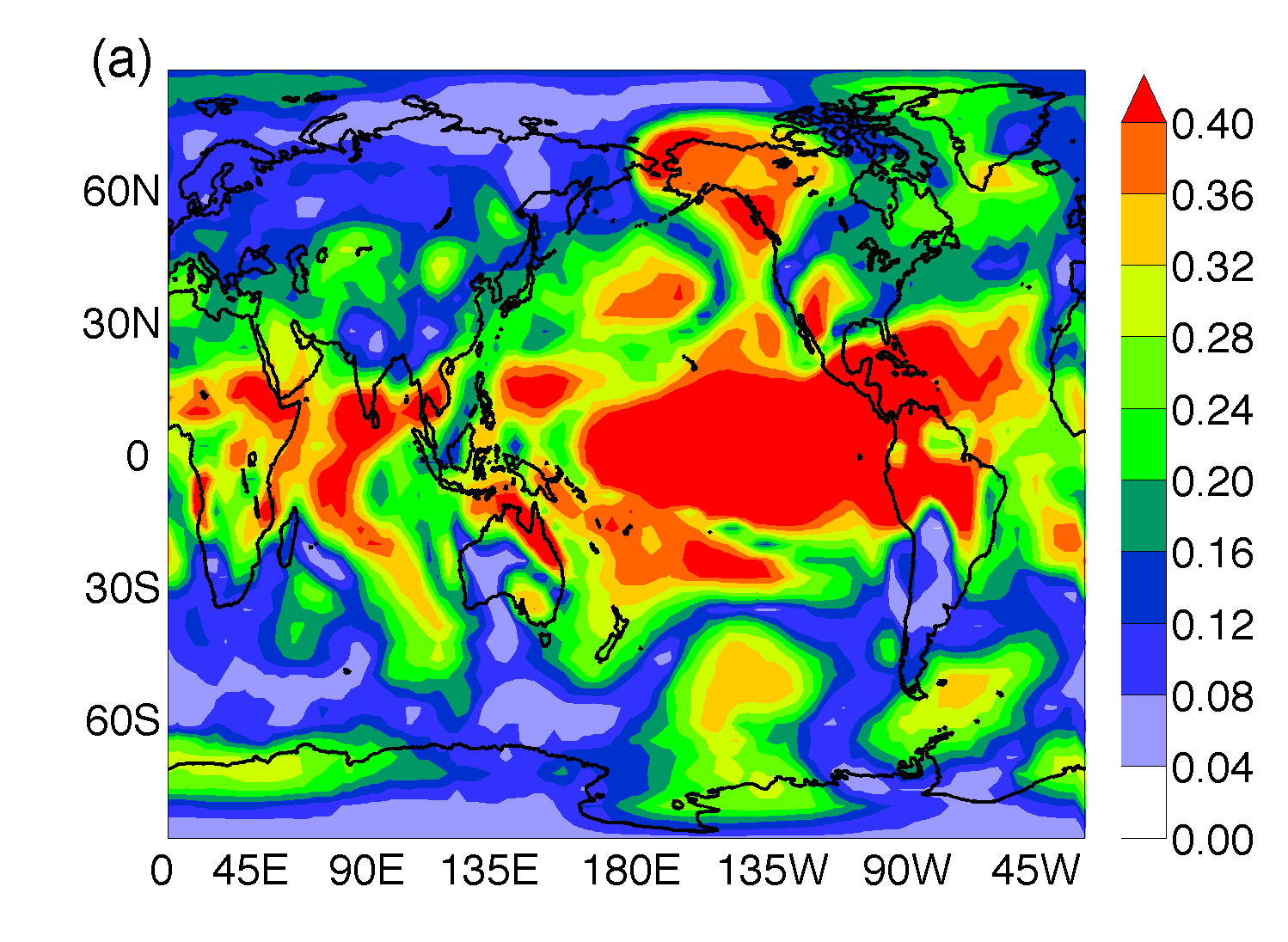}
\includegraphics[width=0.49\textwidth]{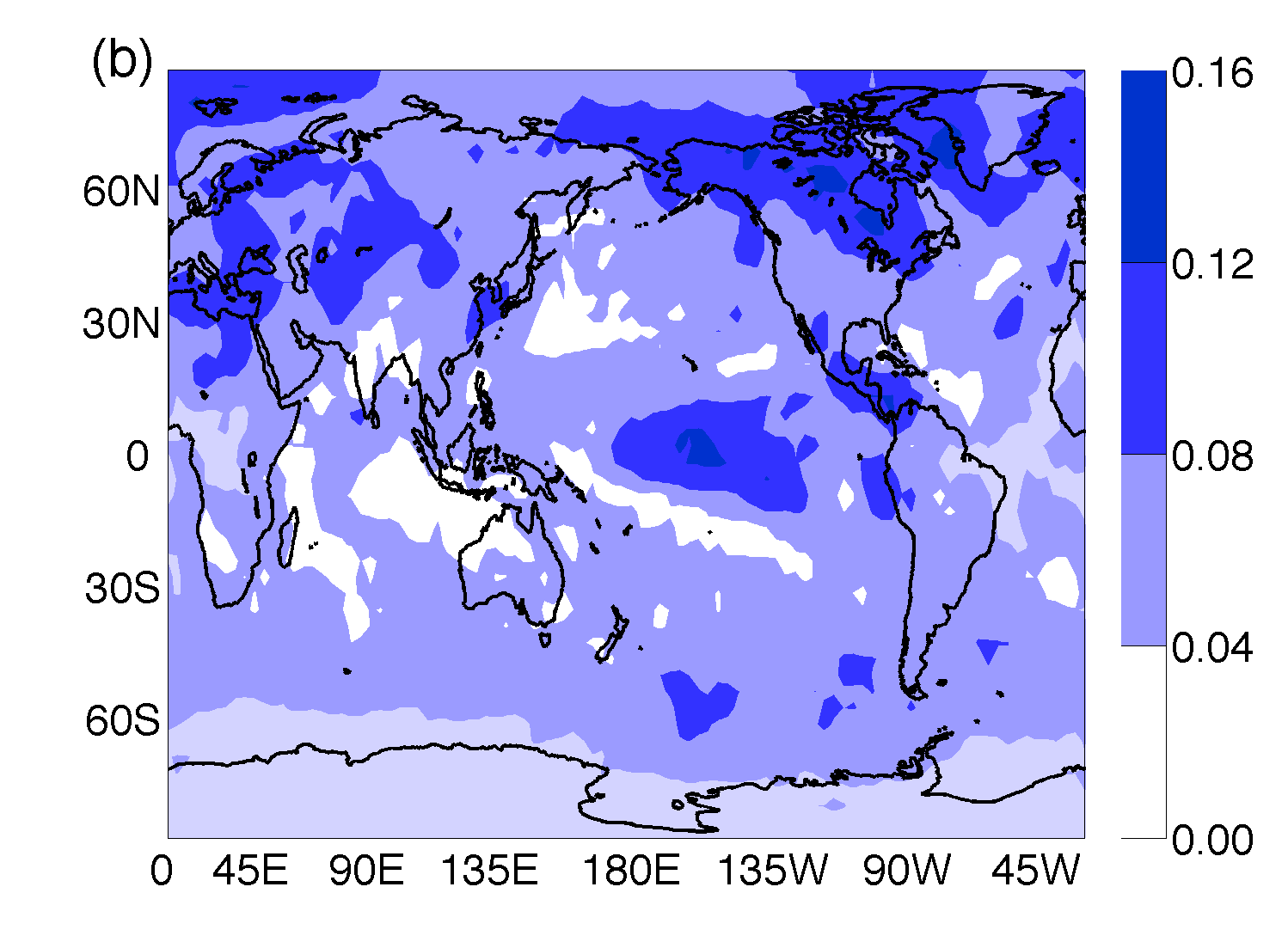}
\includegraphics[width=0.49\textwidth]{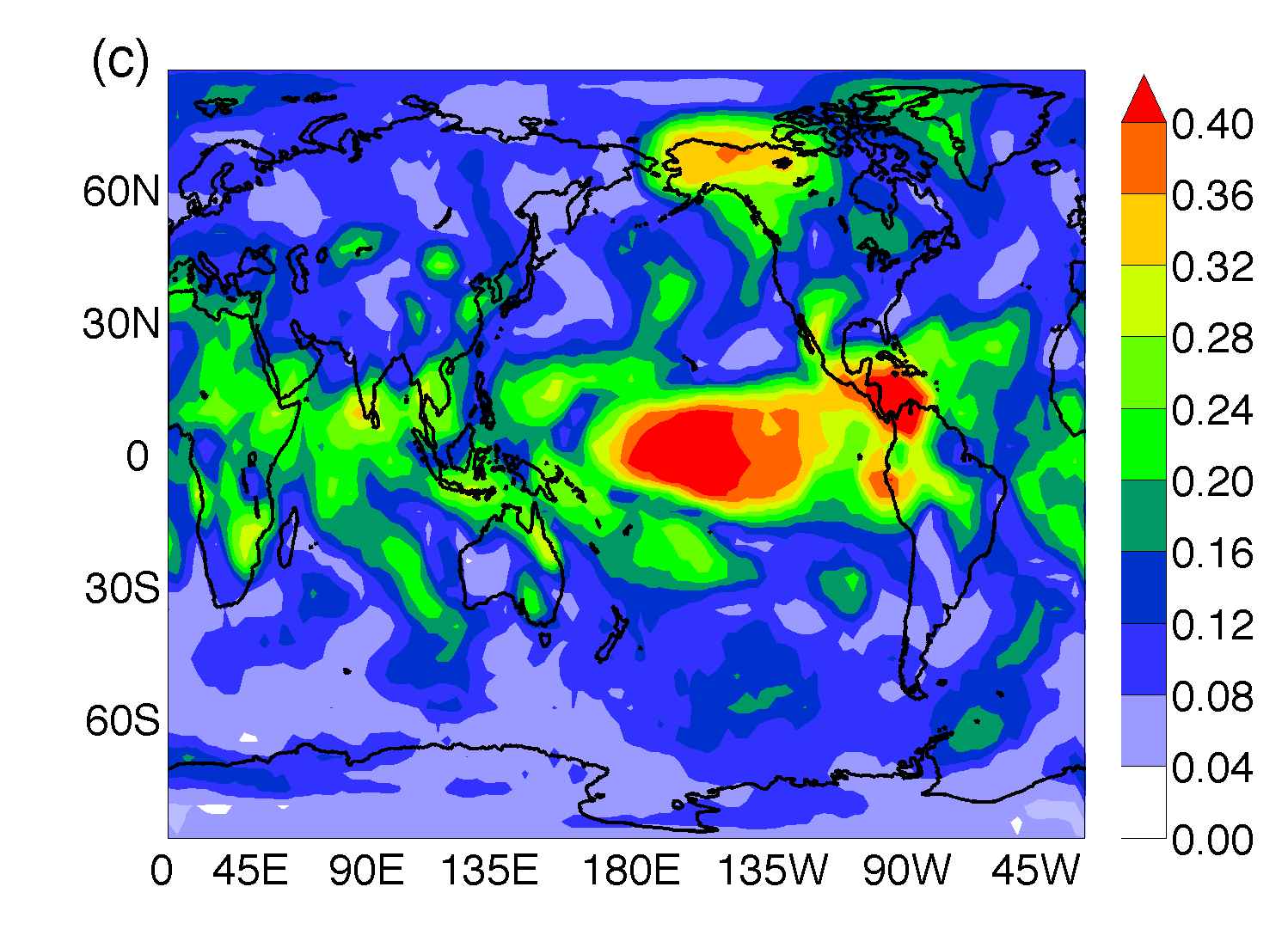}
\includegraphics[width=0.49\textwidth]{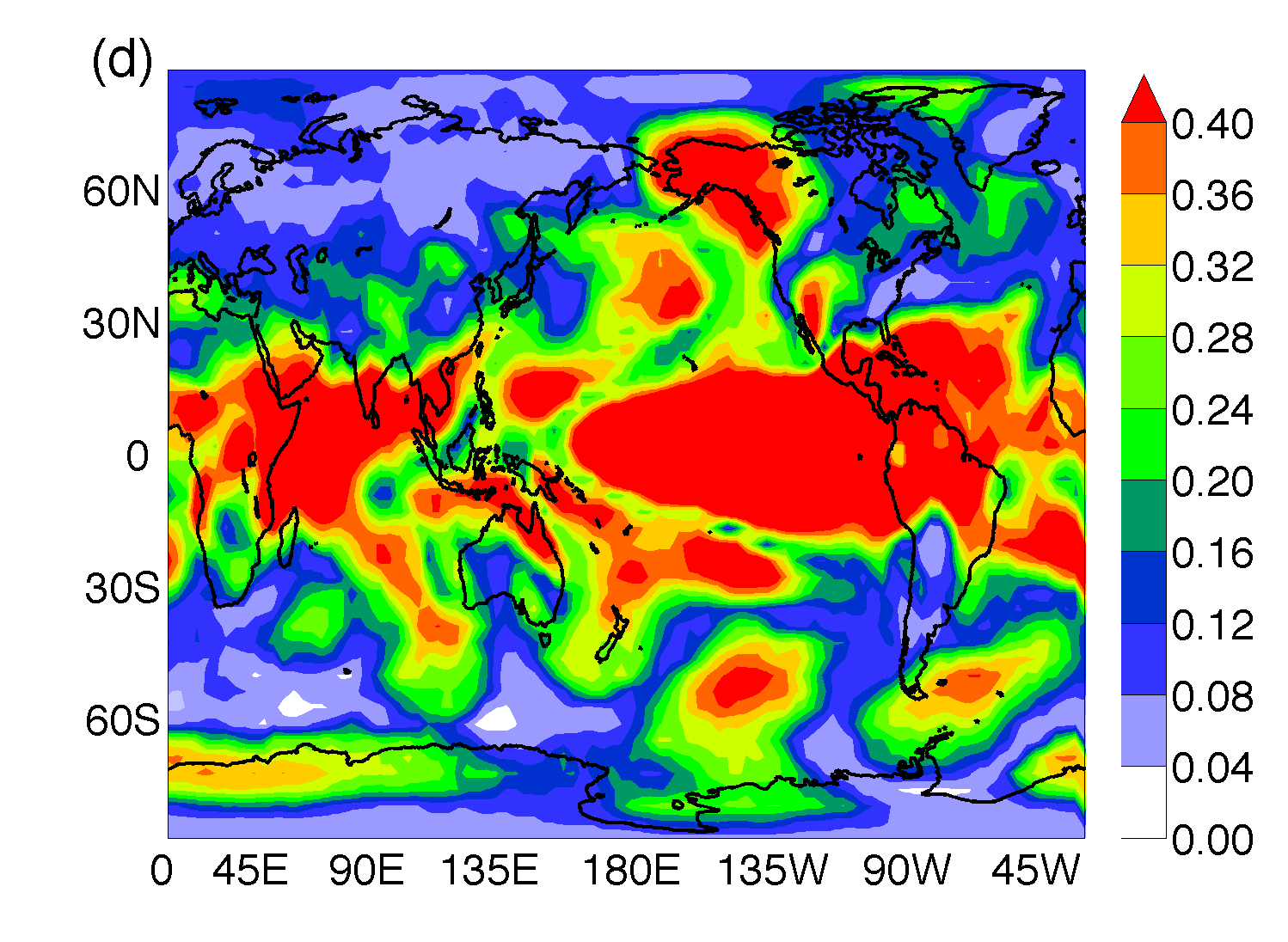}
\caption{Maps of AWC computed from averaged time series, and thus containing information only of the forced component of atmospheric variability. The quantifiers of statistical similarity are as in Fig. \ref{fig:full}: (a) MIH, (b) MI OP intra-season, (c) intra-annual and (d) inter-annual. It can be noticed that in the shorter time scale the tropical area, especially the Pacific ocean has a weak influence, and it grows stronger with increasing time scale. The fact that the maps in panels (a) and (d) are similar suggests that most of the links uncovered by the MIH, panel (a), actually reflect interdependencies in the longer time scale and thus, are seen in panel (d).} \label{fig:awcmean}
\end{center}
\end{figure*}

\begin{figure*}[]
\begin{center}
\includegraphics[width=0.49\textwidth]{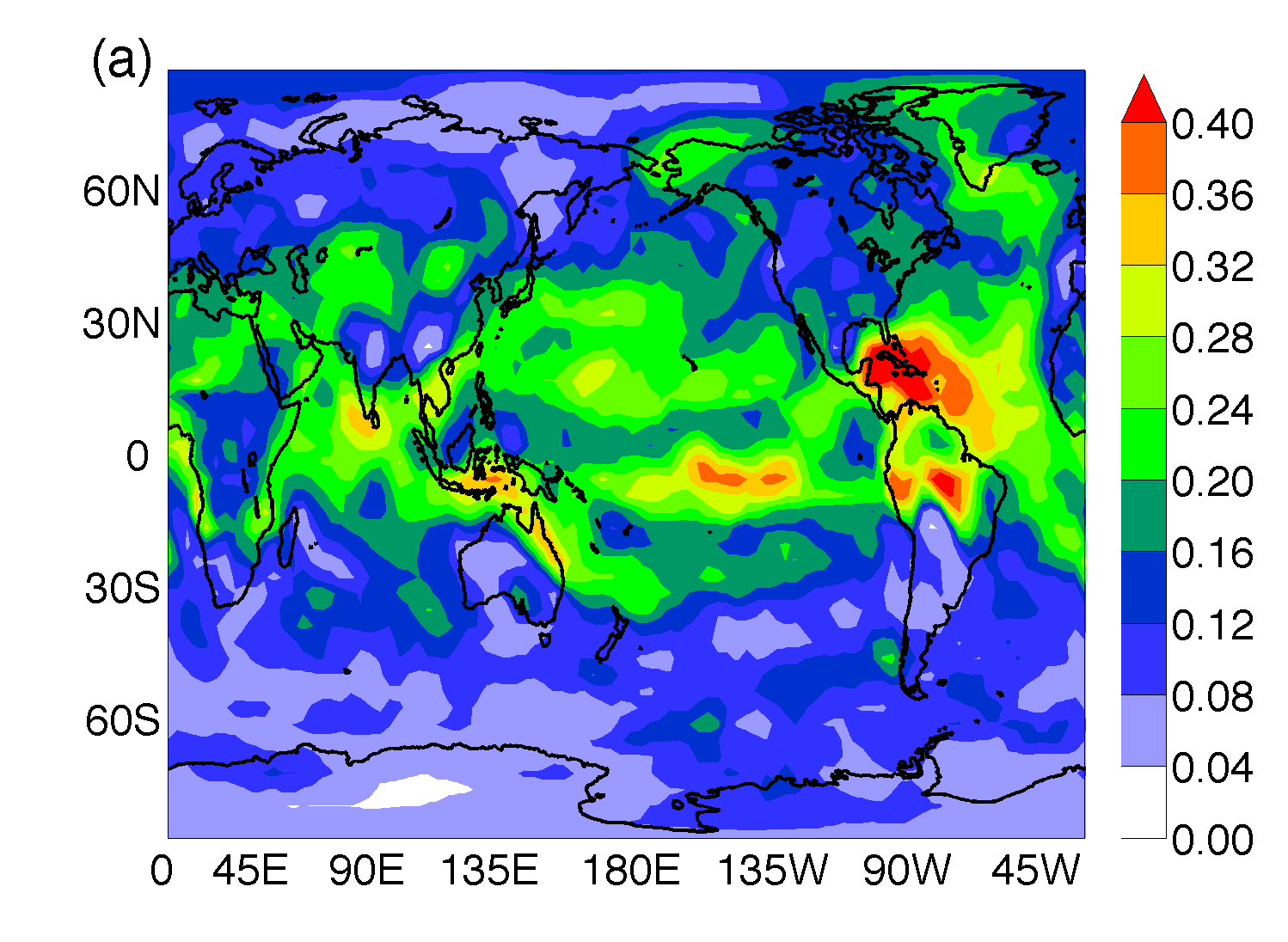}
\includegraphics[width=0.49\textwidth]{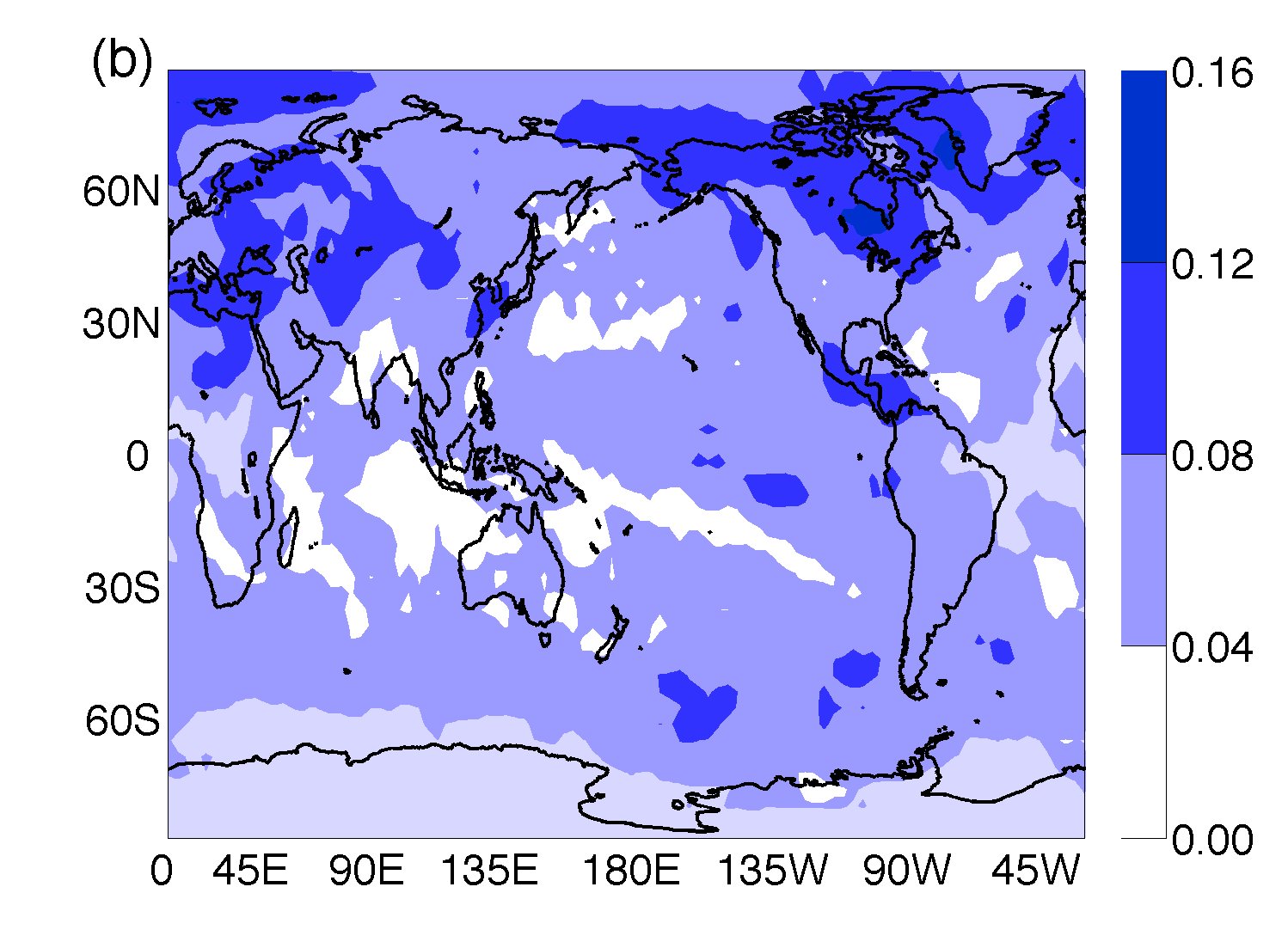}
\includegraphics[width=0.49\textwidth]{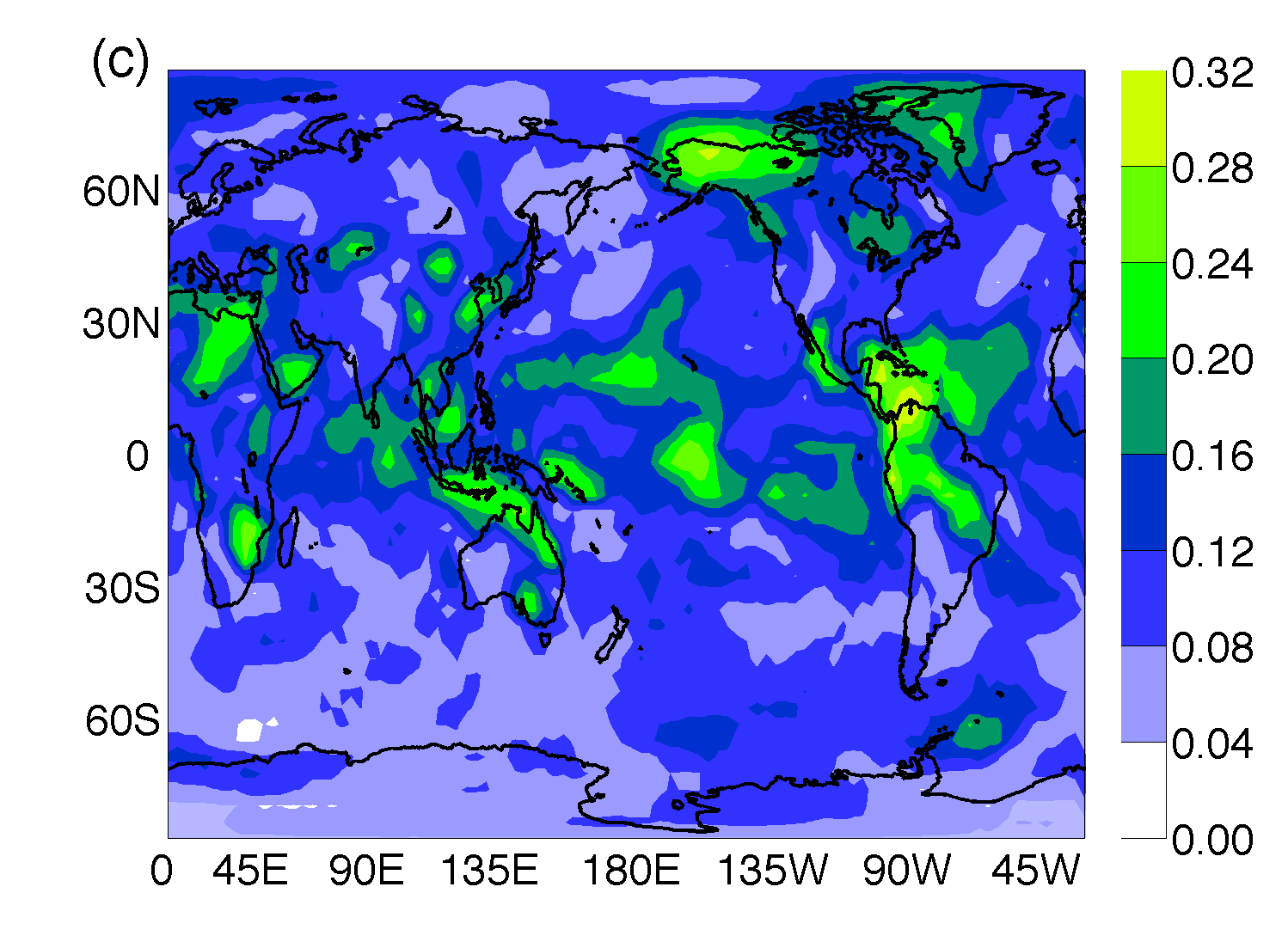}
\includegraphics[width=0.49\textwidth]{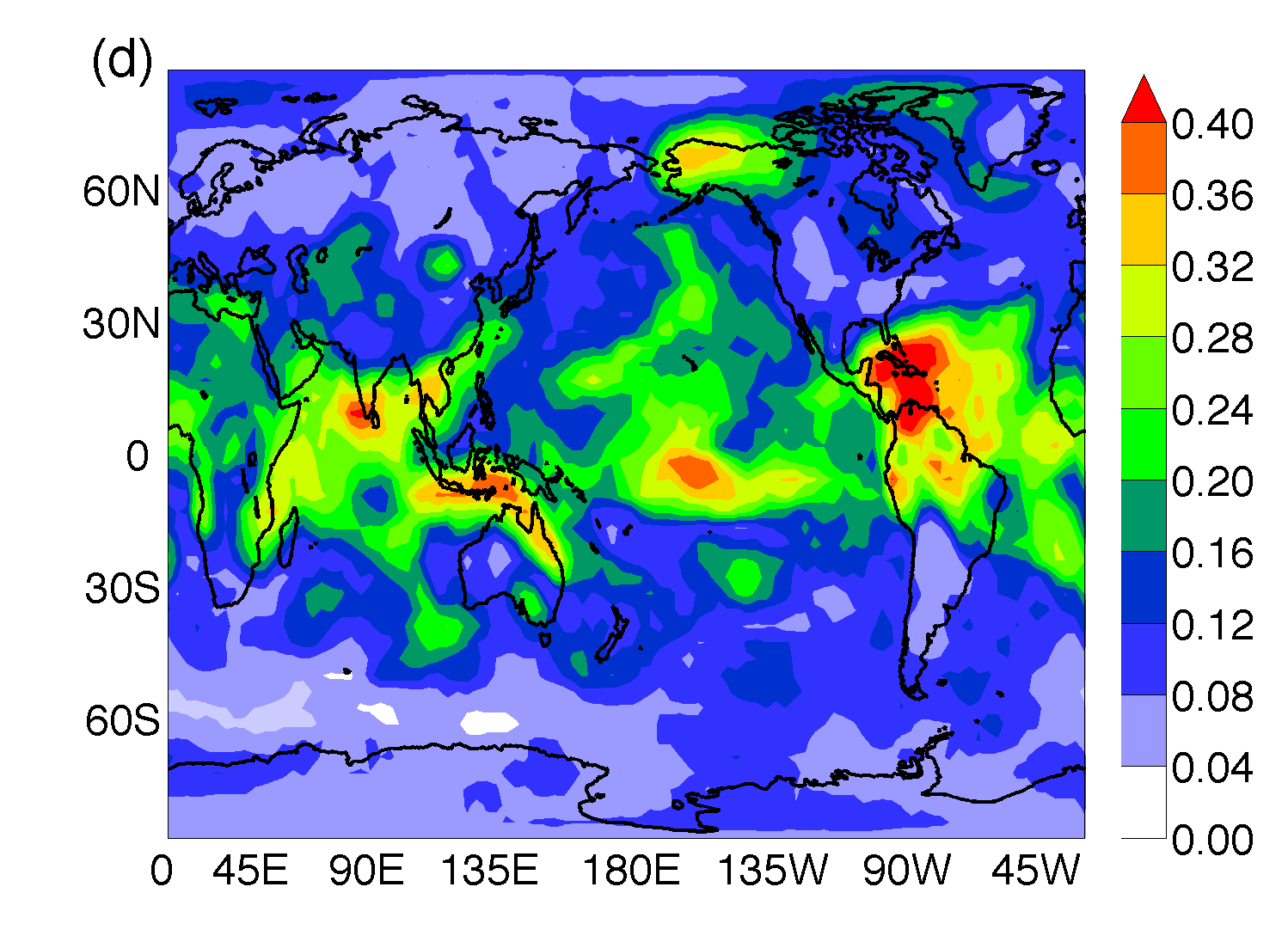}
\caption{Maps of AWC of the forced component of the network when the index ENSO3.4 is removed from the time series (for the description of the index and for the removal procedure, see Sections II and IV). The statistical interdependencies are quantified as in Fig. \ref{fig:full}: (a) MIH, MI OP (b) intra-season, (c) intra-annual and (d) inter-annual. A comparison with Fig. \ref{fig:awcmean} allows assessing the influence of El Ni\~no phenomenon in the network connectivity.} \label{fig:noninomean}
\end{center}
\end{figure*}

\begin{figure*}[]
\begin{center}
\includegraphics[width=0.49\textwidth]{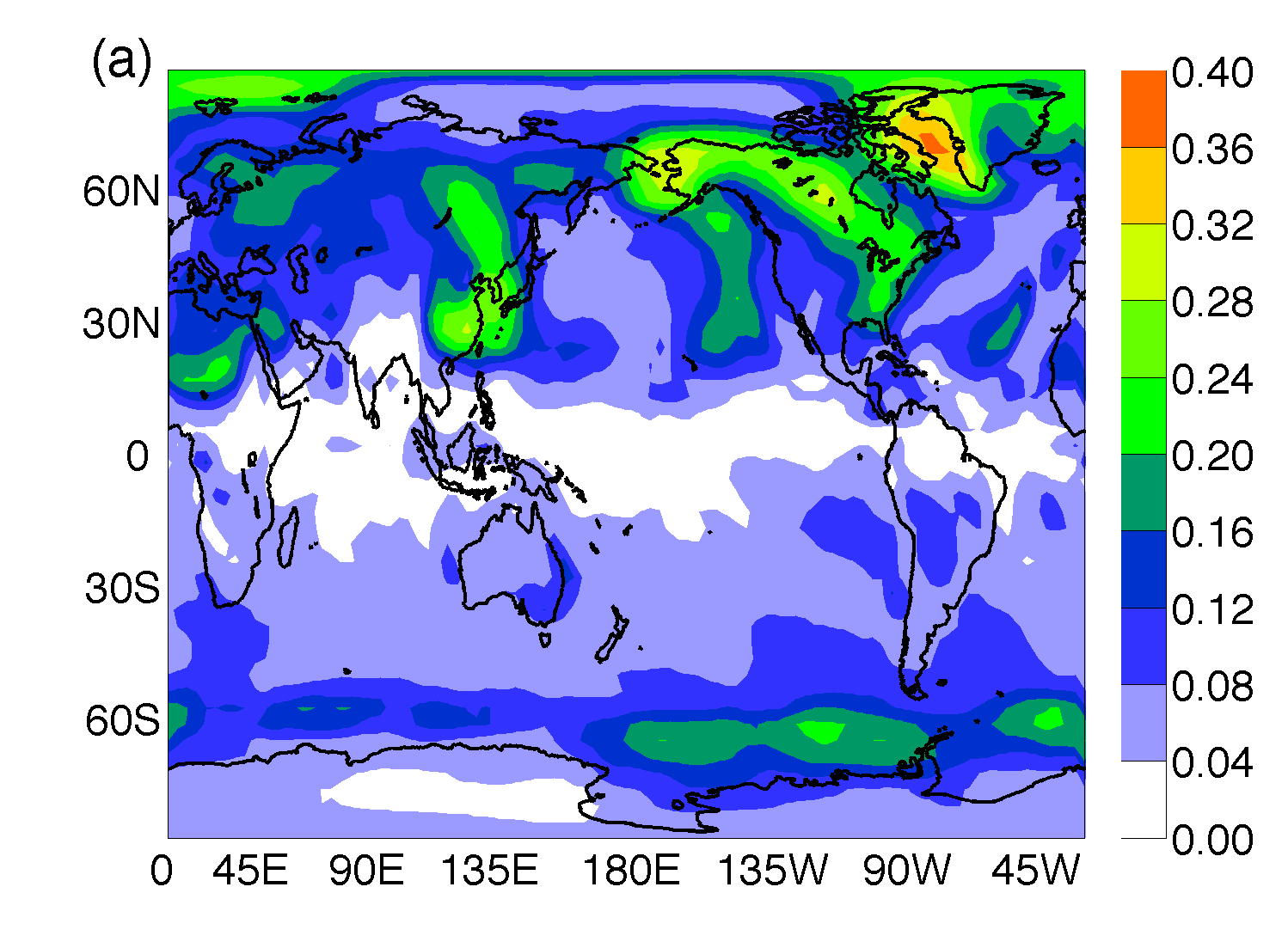}
\includegraphics[width=0.49\textwidth]{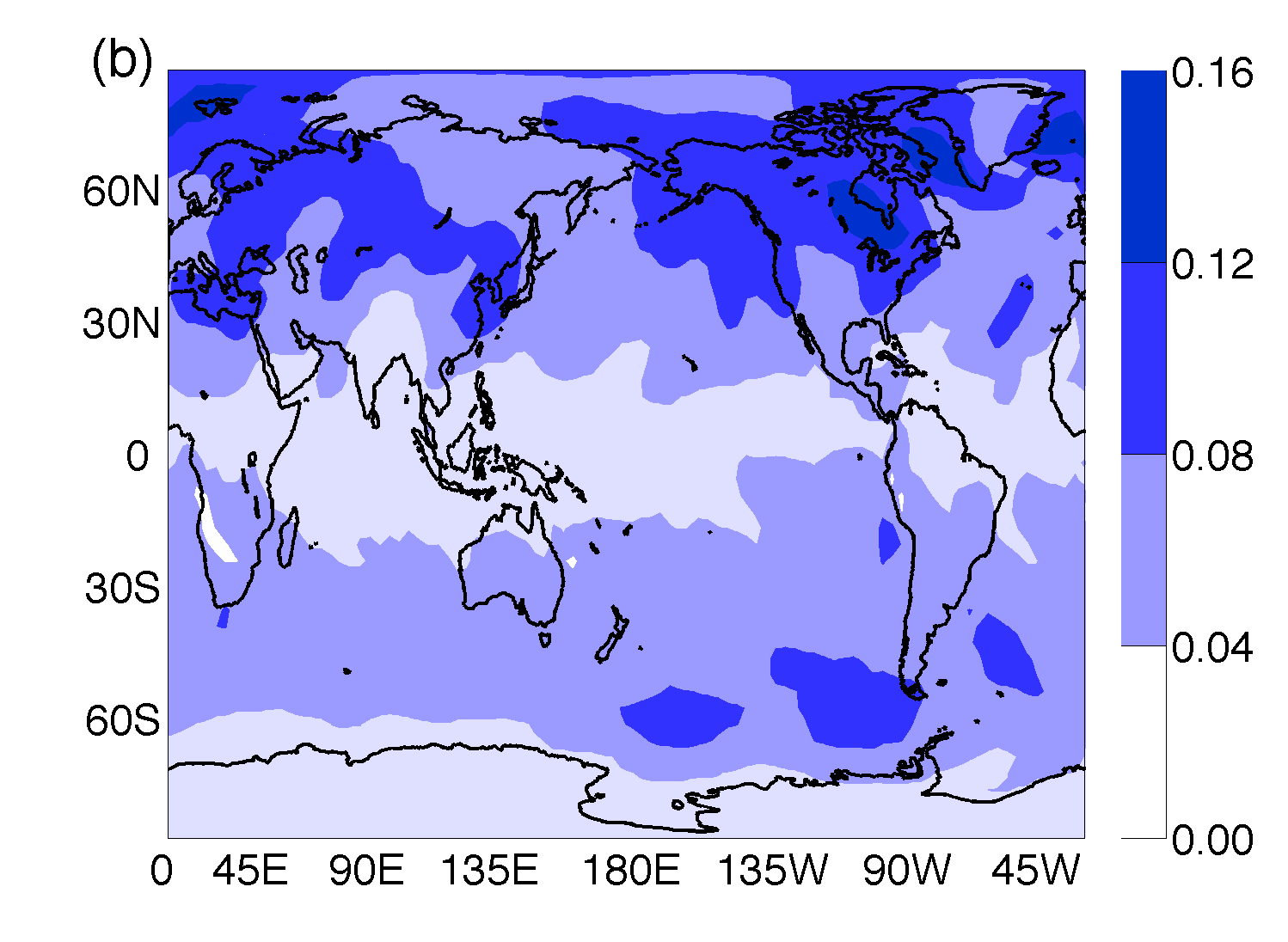}
\includegraphics[width=0.49\textwidth]{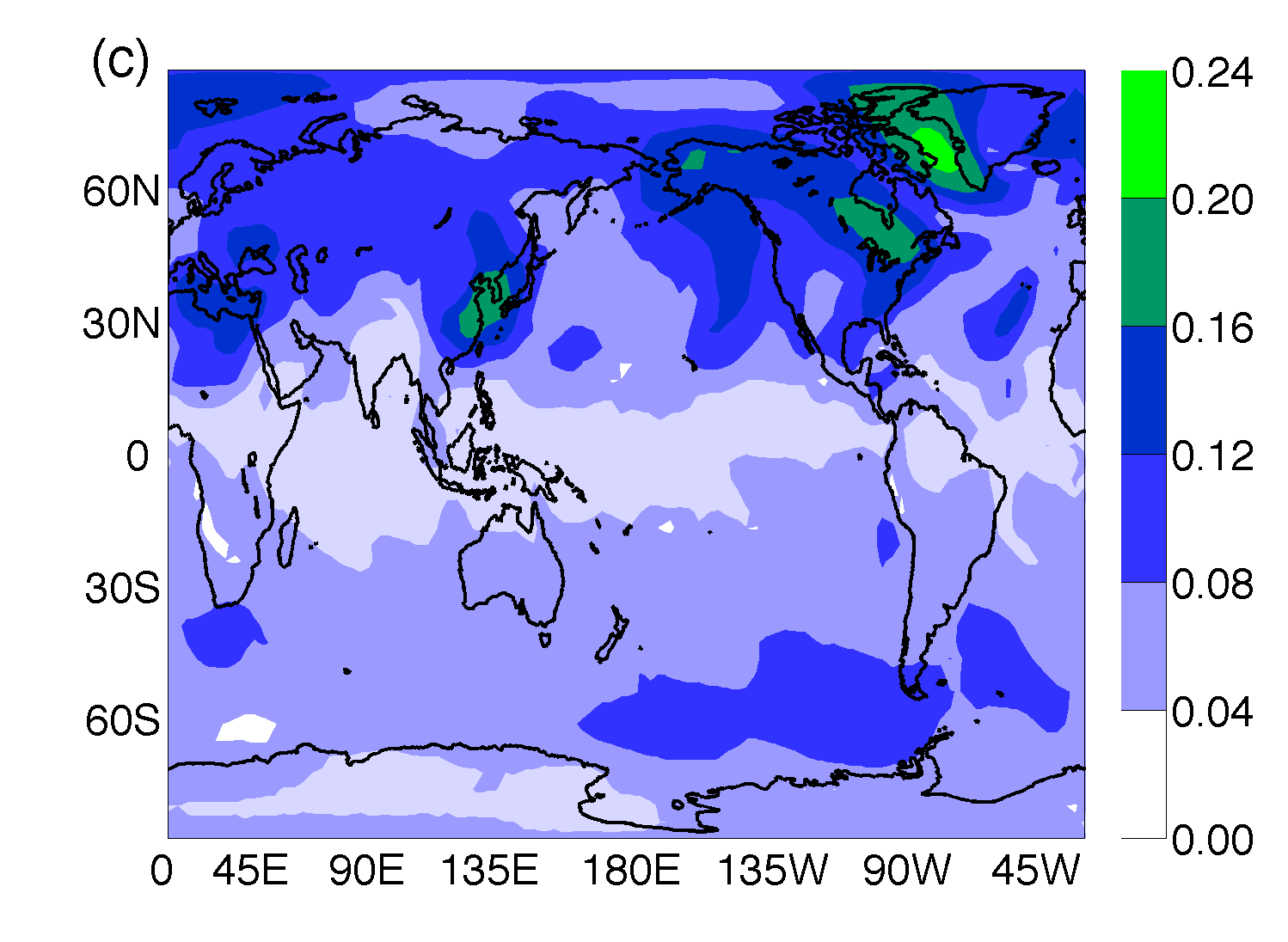}
\includegraphics[width=0.49\textwidth]{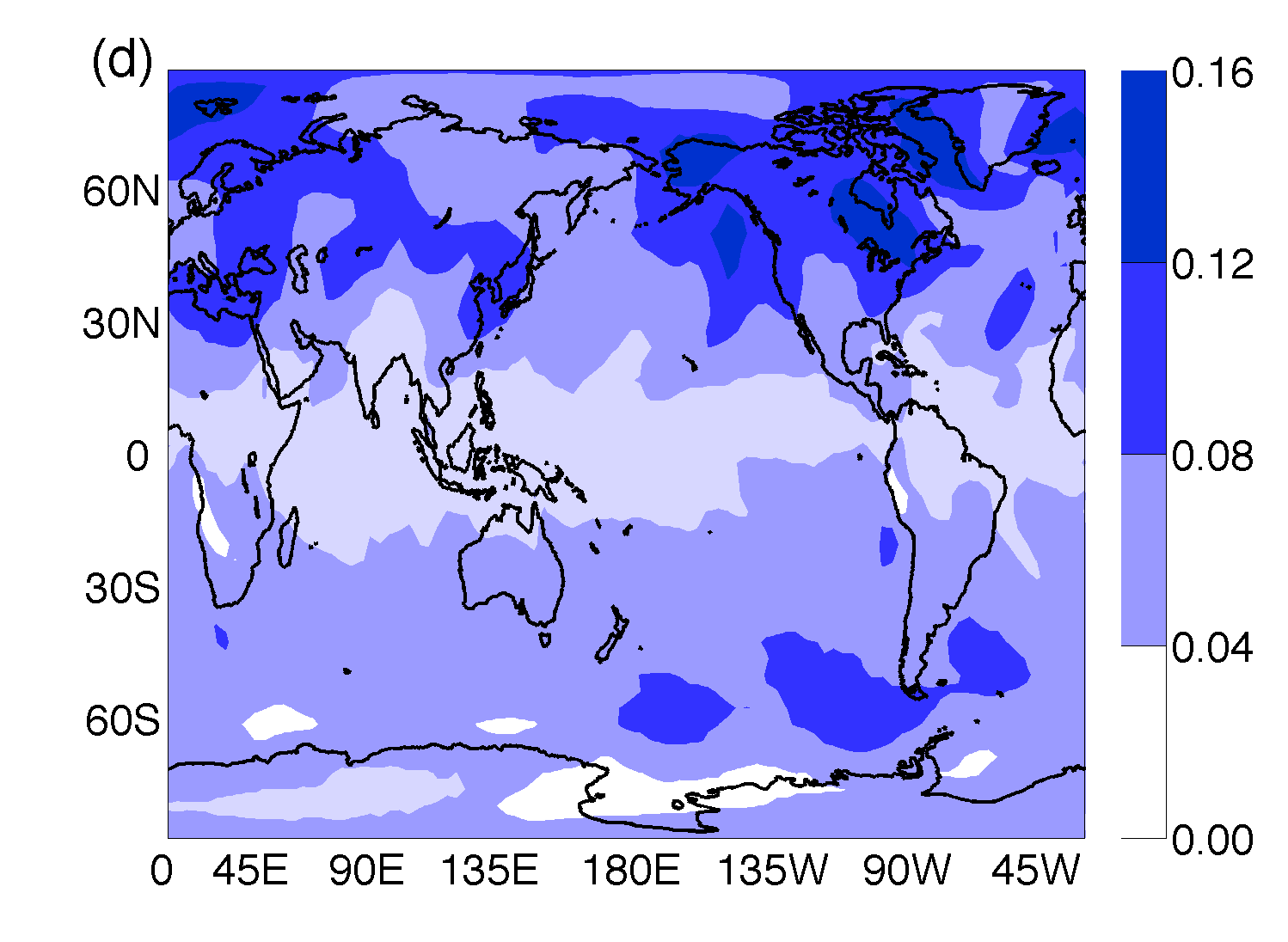}
\caption{Maps of averaged AWC, revealing the internal variability network (see text for details).  The statistical interdependencies are quantified as in Fig. \ref{fig:full} (a) MIH, MI OP (b) intra-season, (c) intra-annual and (d) inter-annual. It can be noticed that in this network the time scale showing more connectivity is the intra-annual time scale. This is consistent with the lower memory of the atmosphere compared with the ocean.} \label{fig:awcmawc}
\end{center}
\end{figure*}

\begin{figure*}[]
\begin{center}
\includegraphics[width=0.49\textwidth]{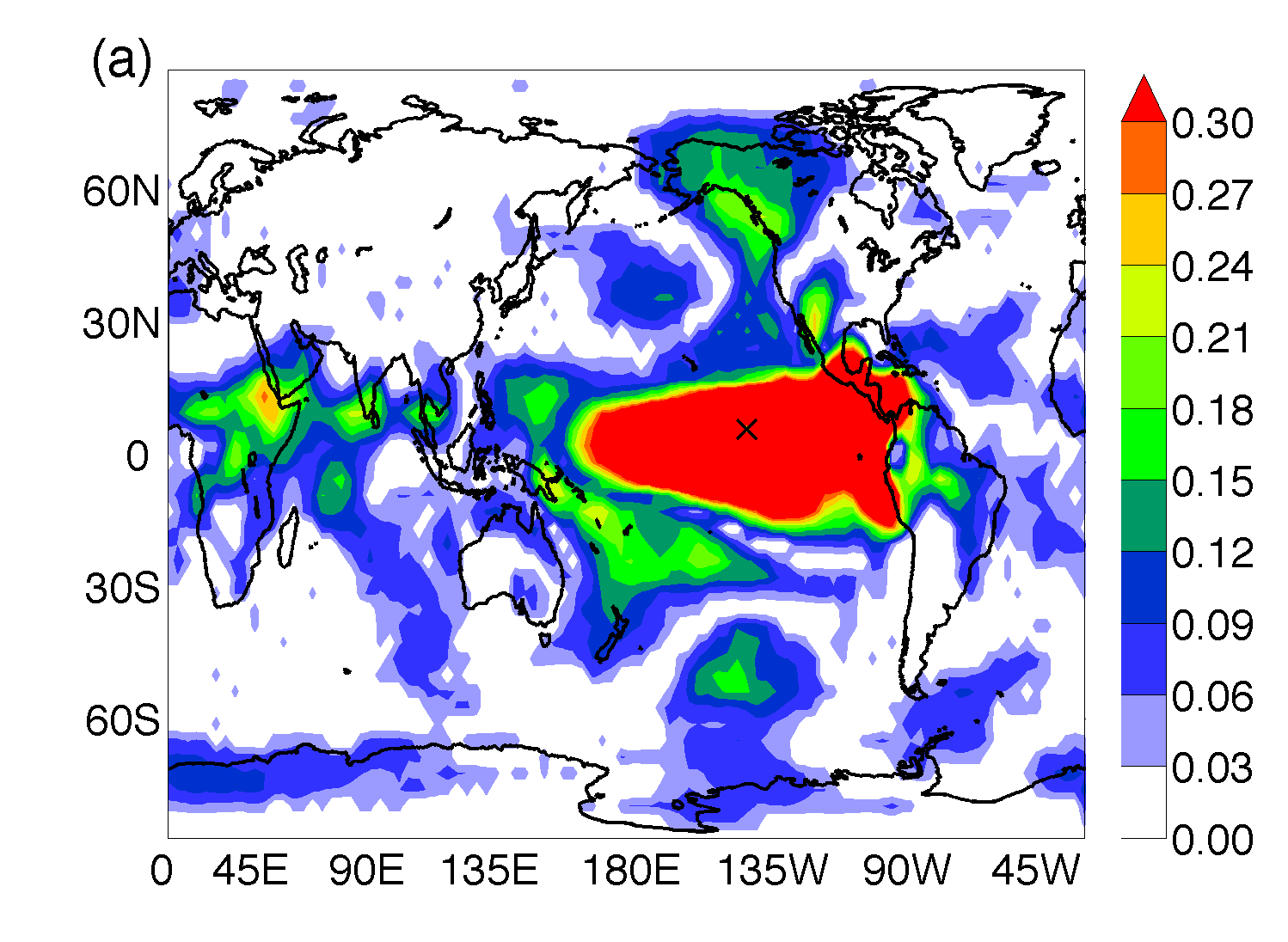}
\includegraphics[width=0.49\textwidth]{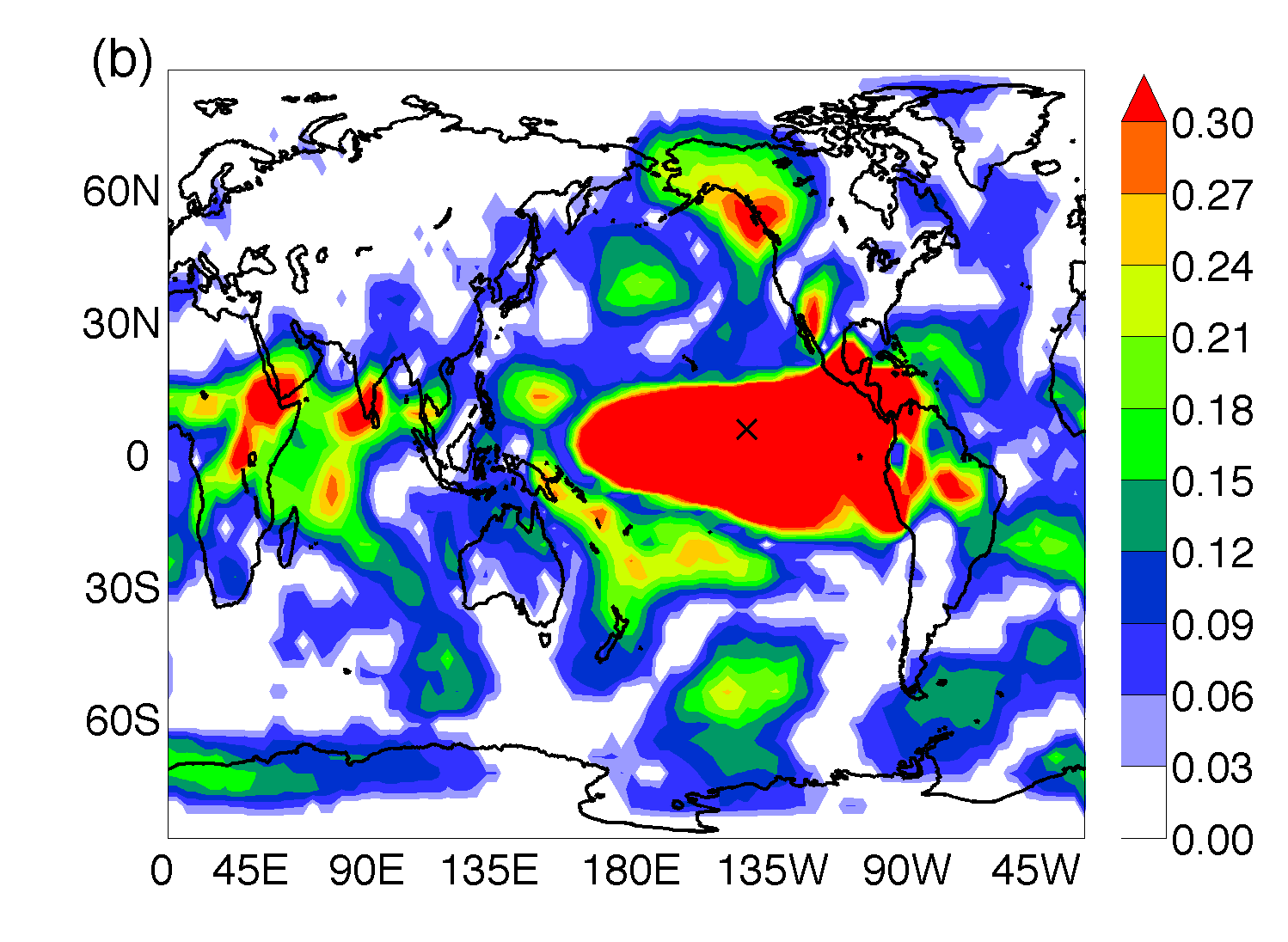}
\includegraphics[width=0.49\textwidth]{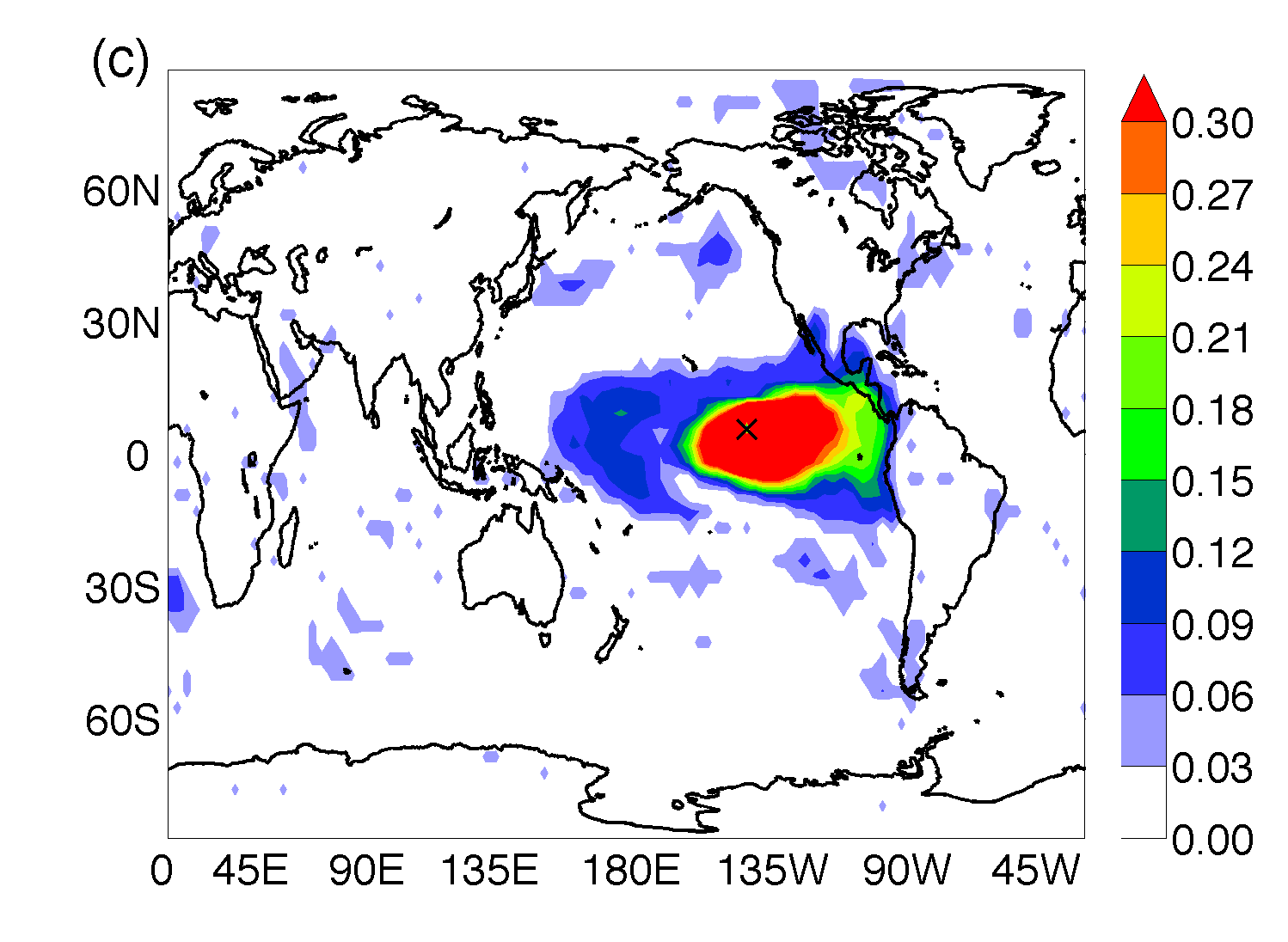}
\includegraphics[width=0.49\textwidth]{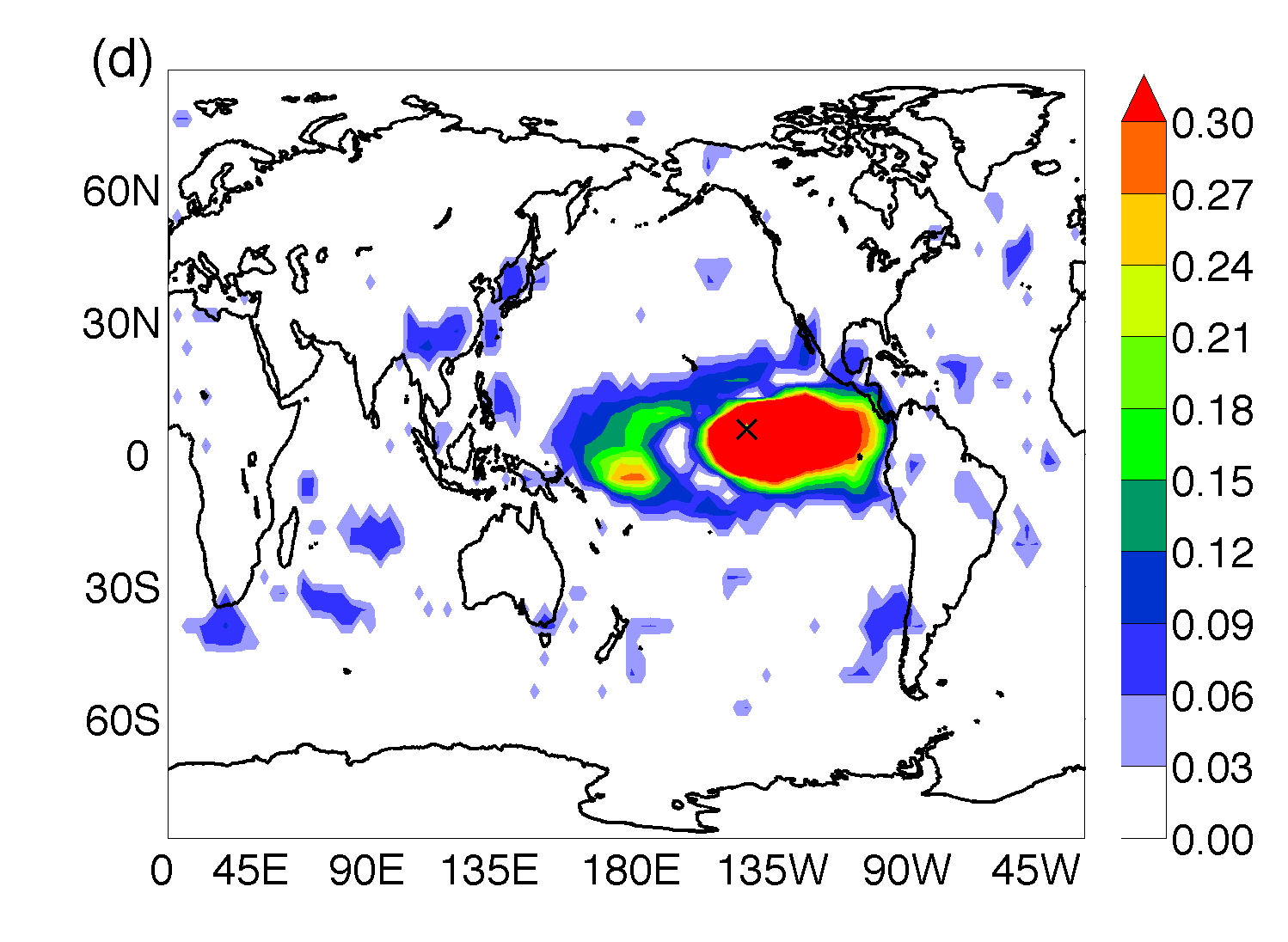}
\caption{Connectivity map of a node in central Pacific (indicated with X). Panels (a) and (b) are computed from forced time series (averaging over nine model realizations); panels (c) and (d) are computed also from forced time series, but with ENSO3.4 linearly removed and thus not containing --to the first order-- contributions due to El Ni\~no . In (a), (c) interdependencies are quantified via MIH; in (b), (d) via MI OP inter-annual time scale.} \label{fig:ninomean}
\end{center}
\end{figure*}

\begin{figure*}[]
\begin{center}
\includegraphics[width=0.49\textwidth]{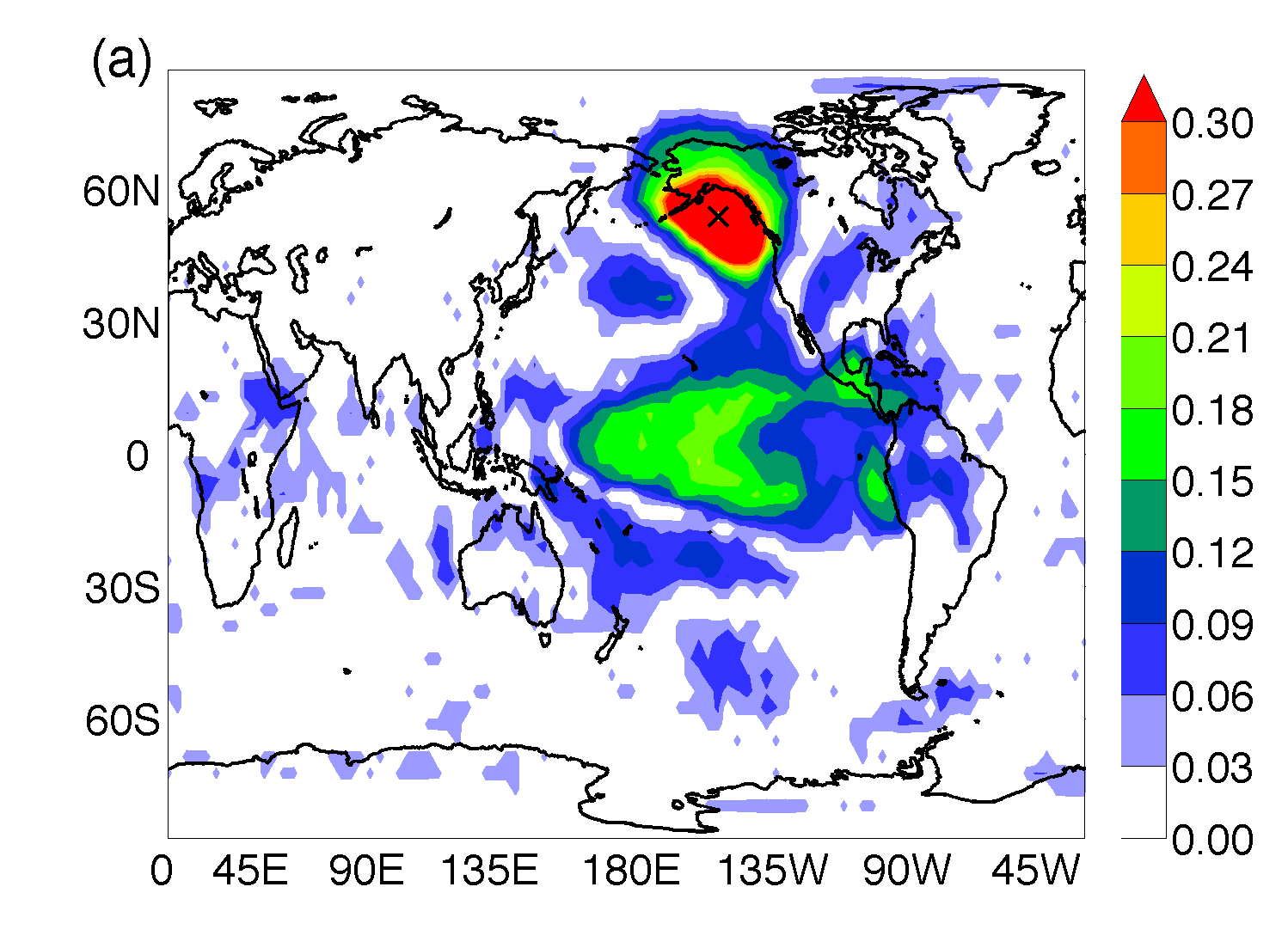}
\includegraphics[width=0.49\textwidth]{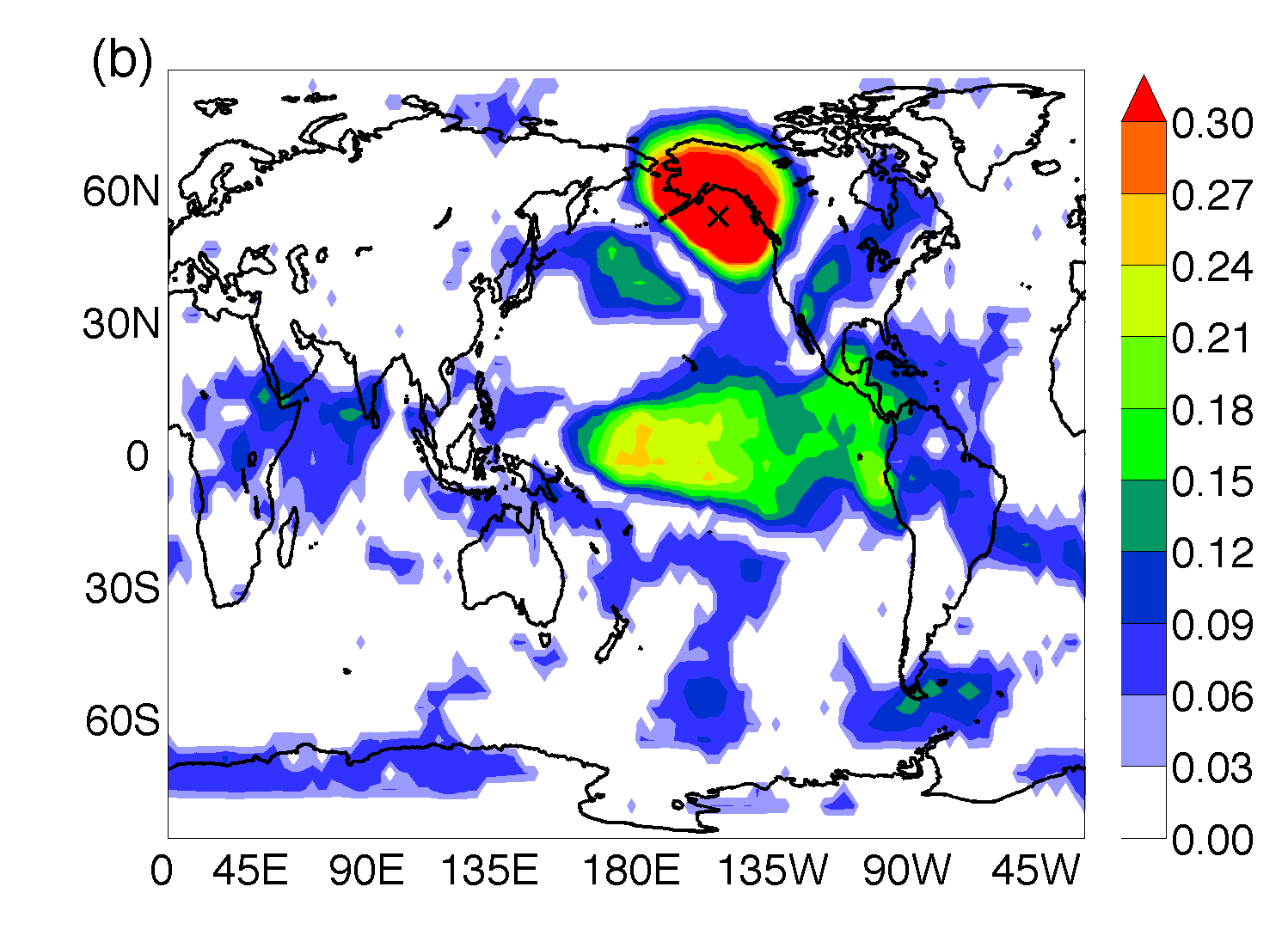}
\includegraphics[width=0.49\textwidth]{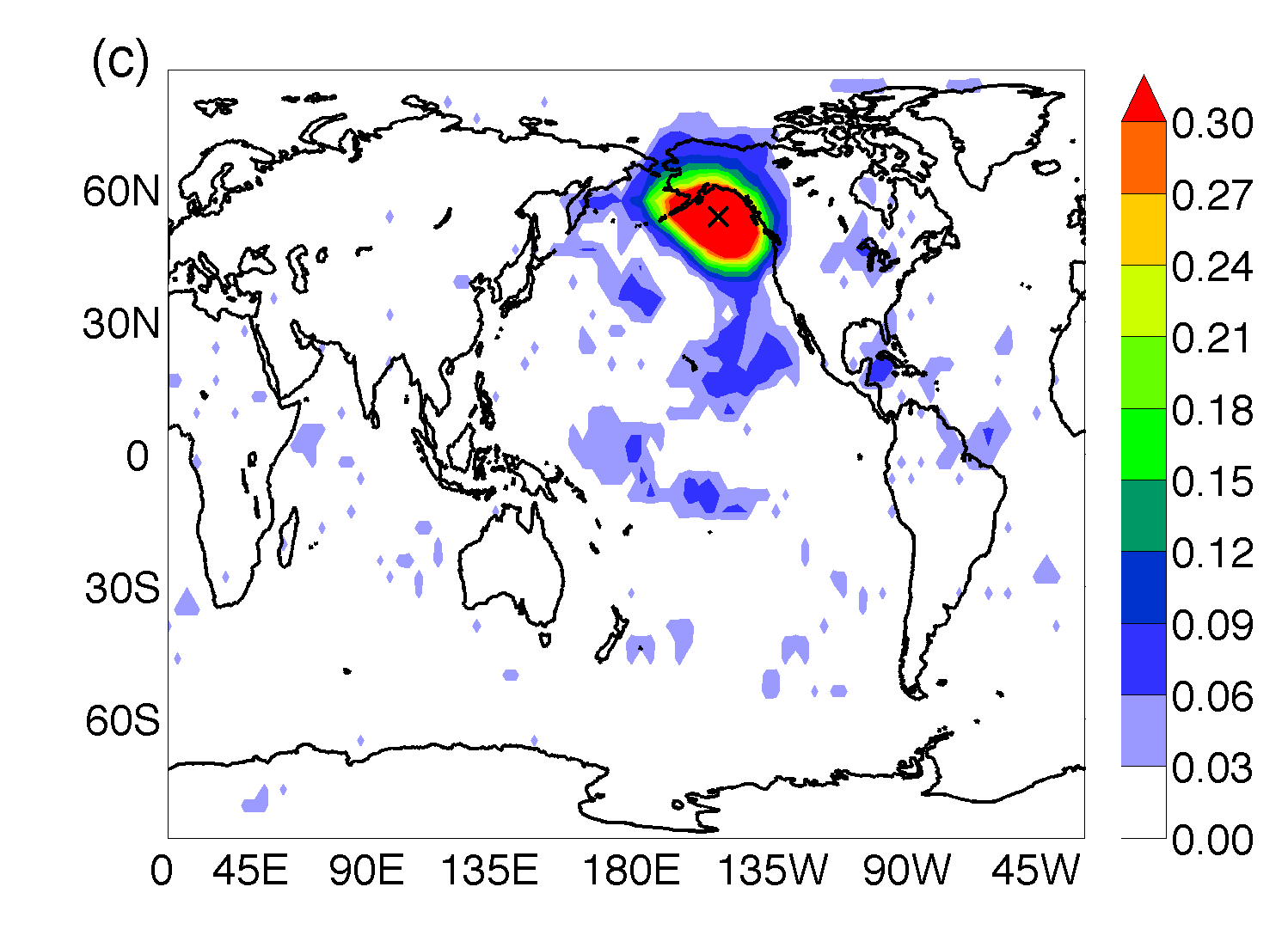}
\includegraphics[width=0.49\textwidth]{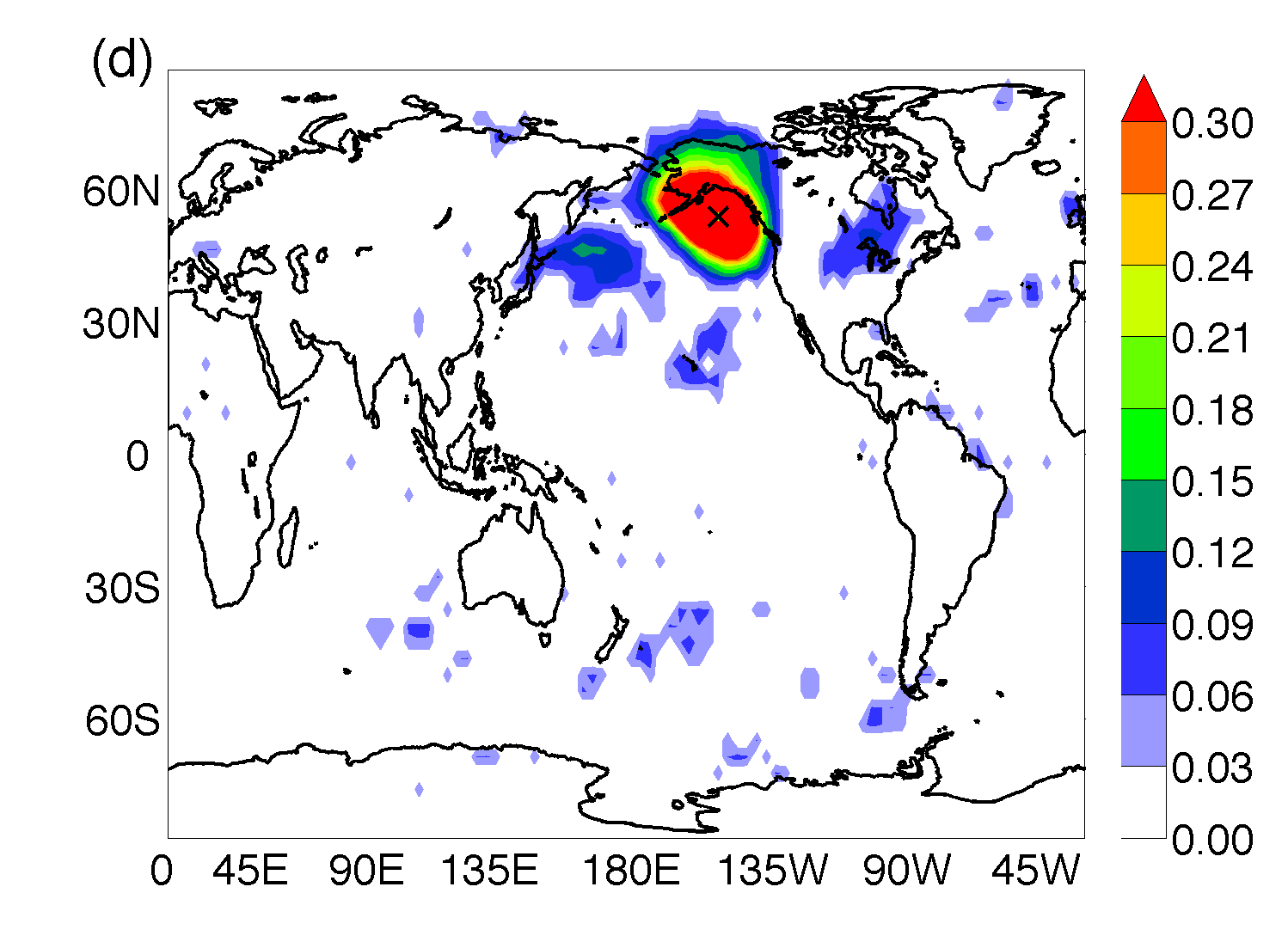}
\caption{As Fig. \ref{fig:ninomean} but of a node near Alaska (indicated with X). Comparing with Fig. \ref{fig:ninomean} one can notice that the teleconnection between this region and the Pacific in due mainly to El Ni\~no.} \label{fig:alaskamean}
\end{center}
\end{figure*}

\begin{figure*}[]
\begin{center}
\includegraphics[width=0.49\textwidth]{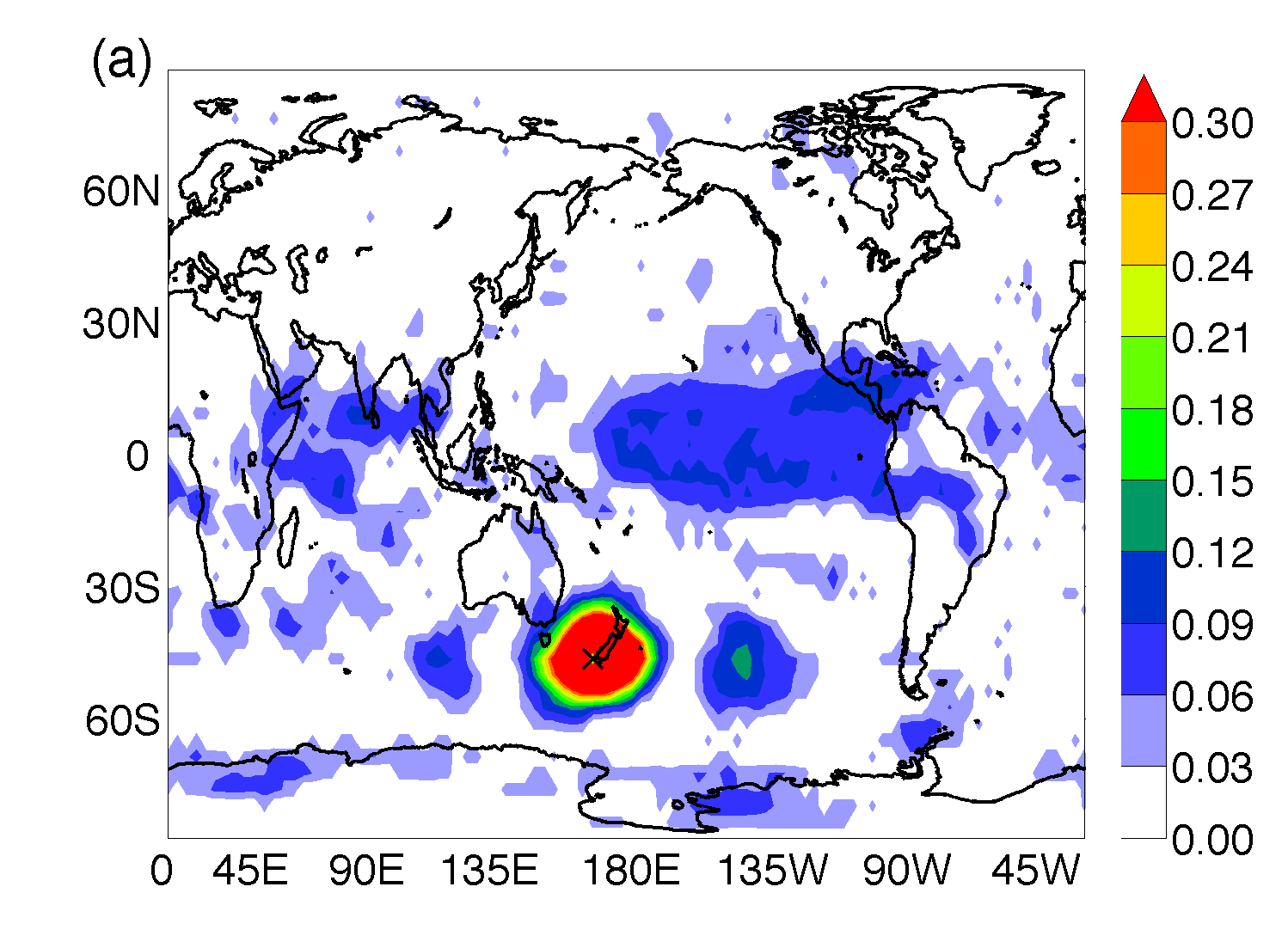}
\includegraphics[width=0.49\textwidth]{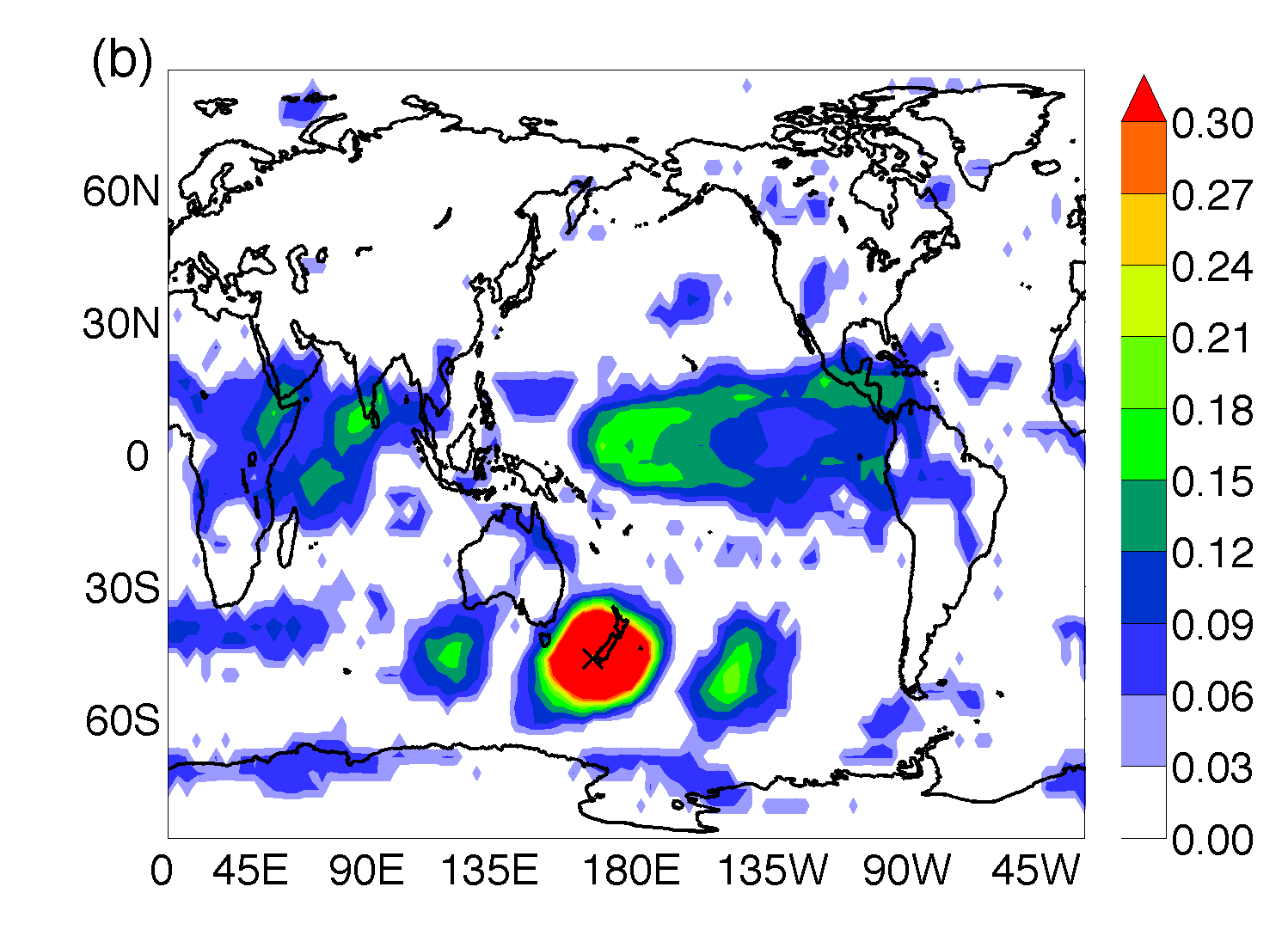}
\includegraphics[width=0.49\textwidth]{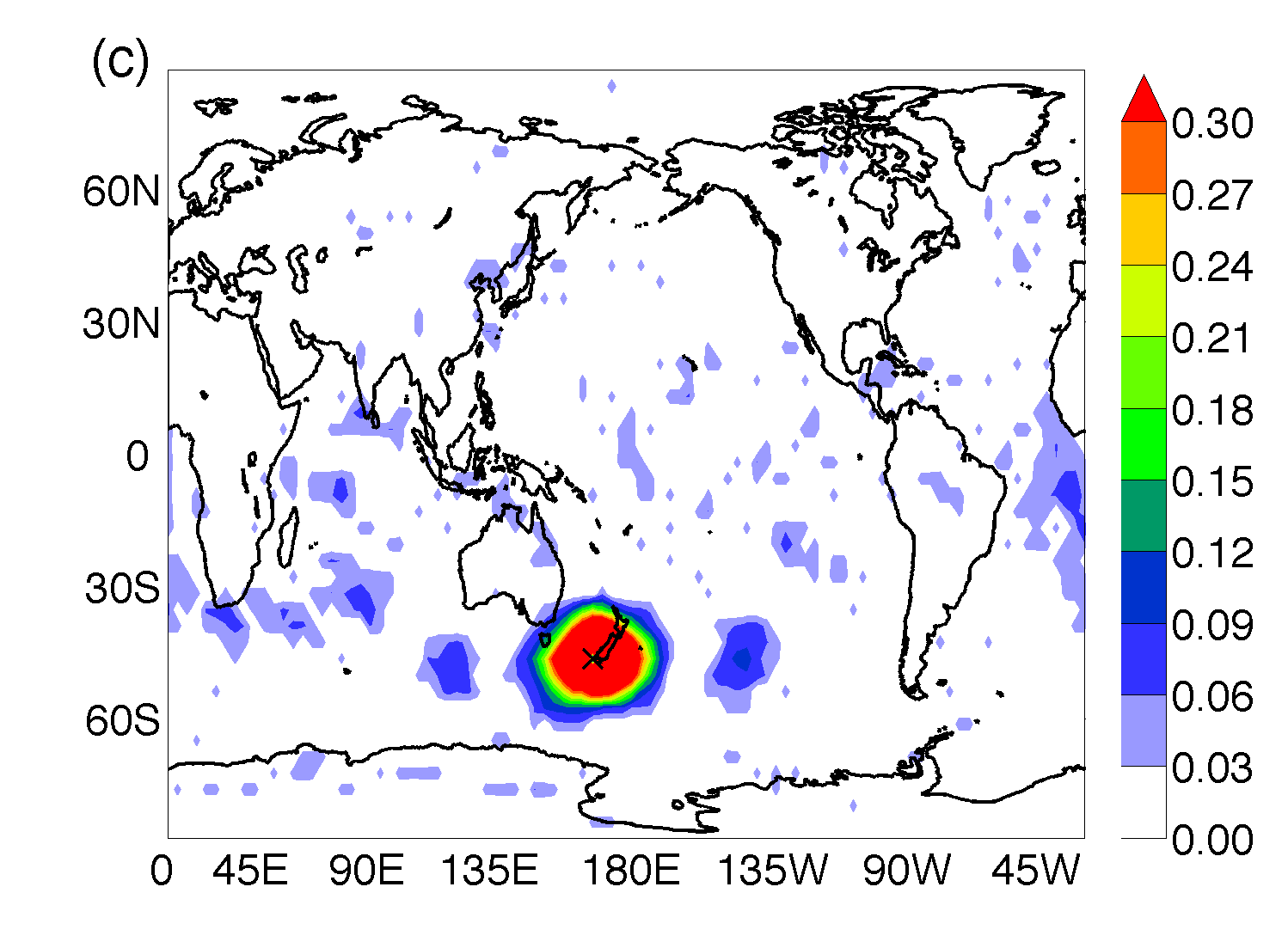}
\includegraphics[width=0.49\textwidth]{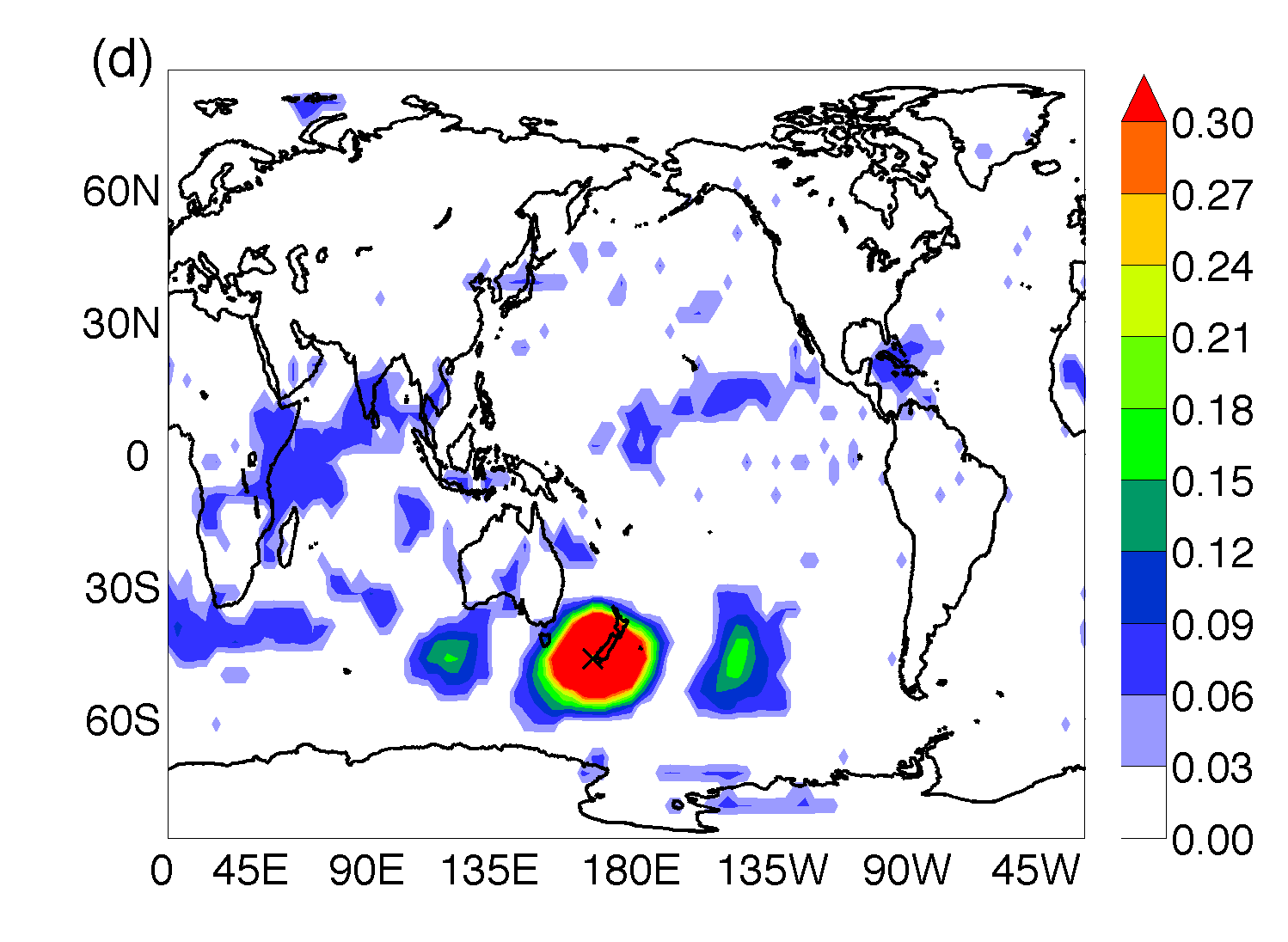}
\caption{As Fig. \ref{fig:ninomean} but of a node near New Zealand (indicated with X). In panels (b) and (d) the MI OP is tuned to inter-annual time scale. Compare with Figs \ref{fig:ninomean} and \ref{fig:alaskamean}.} \label{fig:newZmean}
\end{center}
\end{figure*}

\begin{figure*}[]
\begin{center}
\includegraphics[width=0.49\textwidth]{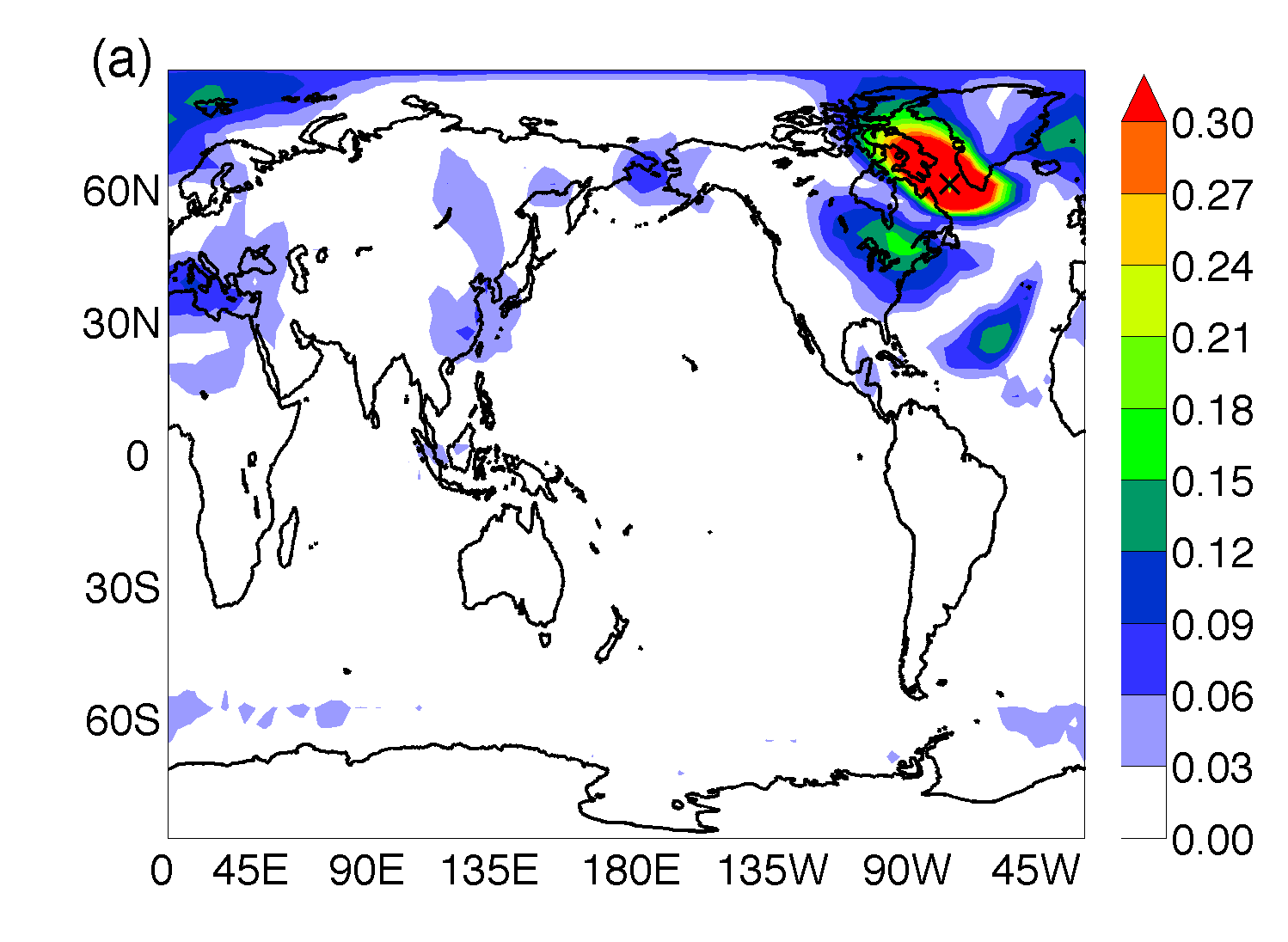}
\includegraphics[width=0.49\textwidth]{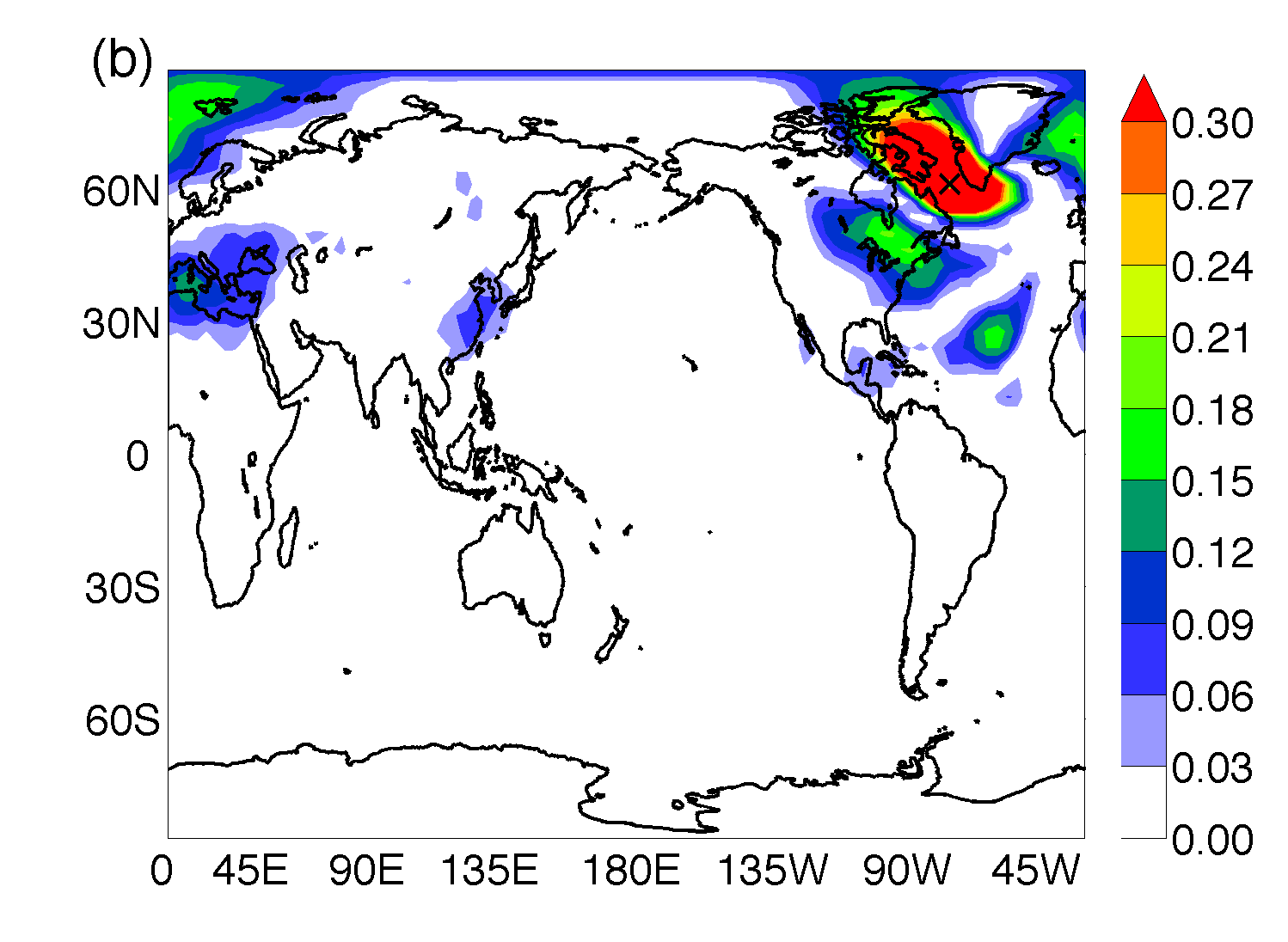}
\includegraphics[width=0.49\textwidth]{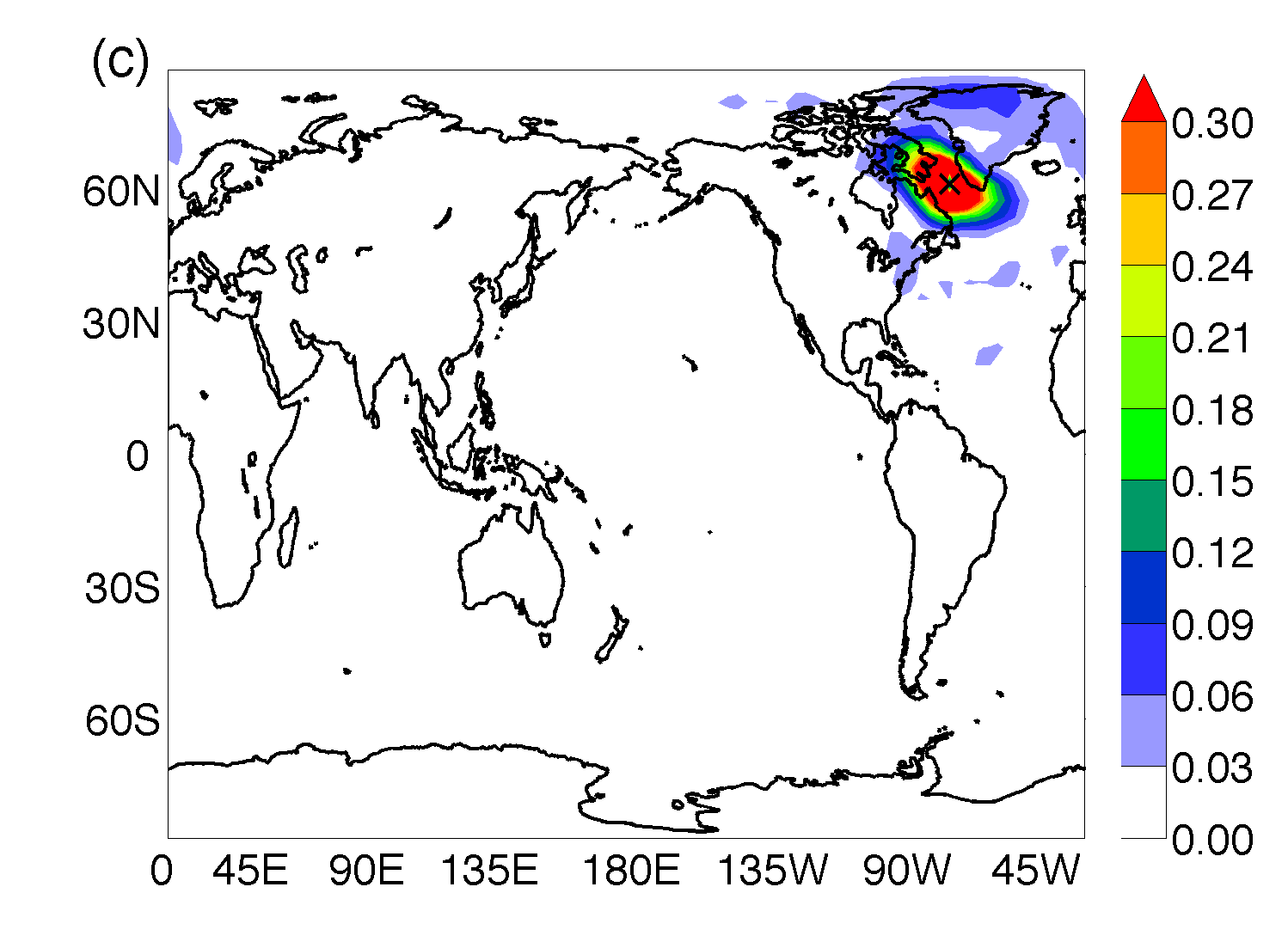}
\includegraphics[width=0.49\textwidth]{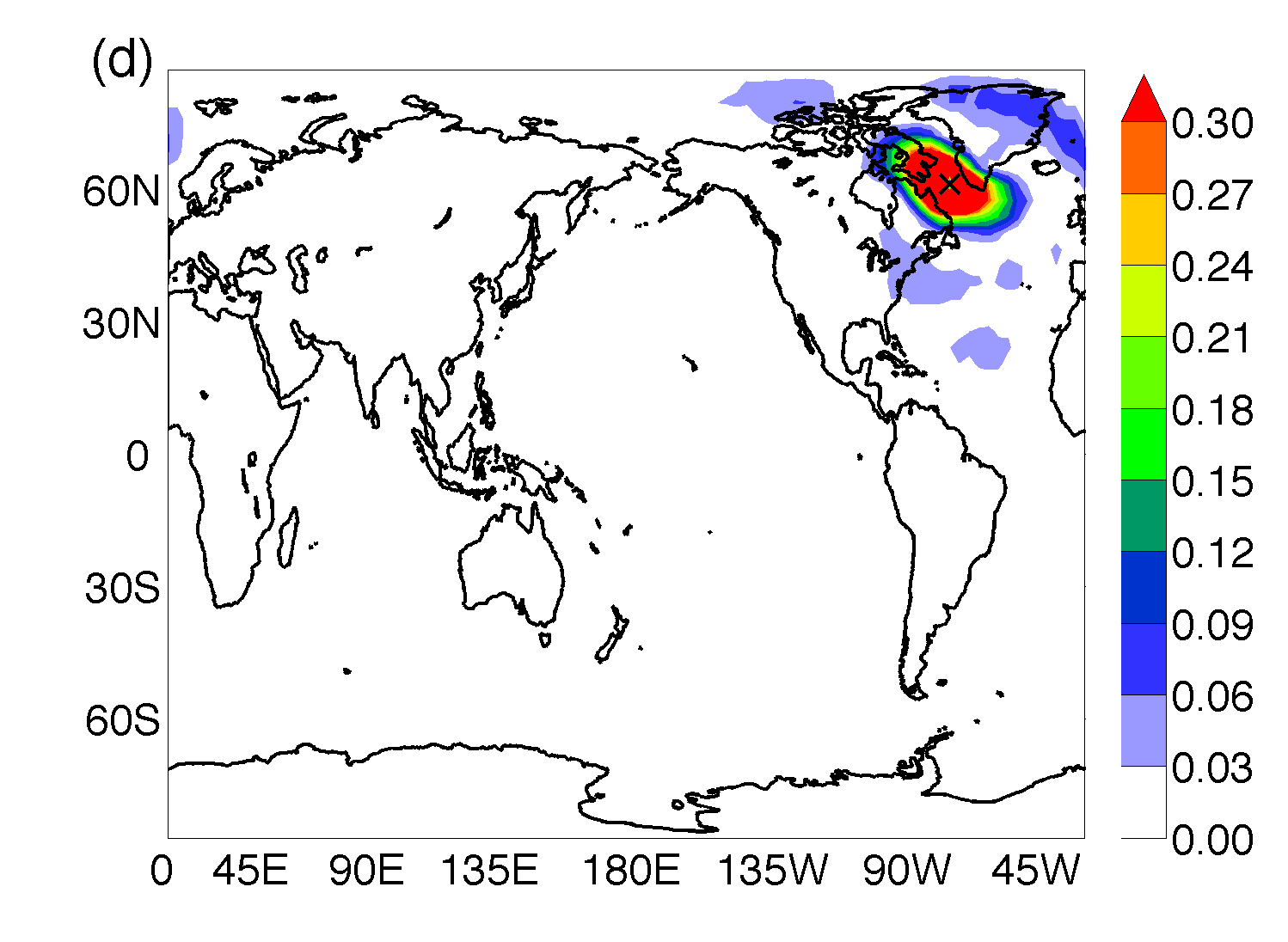}
\caption{Maps of internal variability showing the connectivity of a node in the Labrador Sea (indicated with X). Panel (a), (b) correspond to the original internal time series as in fig \ref{fig:awcmawc}; in panels (c), (d) the NAO was linearly removed and thus the links do not contain --to the first order-- contributions due to the North Atlantic Oscillation. In (a), (c) interdependencies are quantified via MIH; in (b), (d) via MI OP intra-annual time scale.} \label{fig:labramawc}
\end{center}
\end{figure*}

 \subsection{Model Validation}
While the ICTP-AGCM model has been used extensively in the literature (see, e.g. \citep{KucharskiInternal,Kucharski2005Decadal,Molteni2003Atmospheric,barreiro2009Influence} and references therein), the model has not yet been validated in the context of climate networks. Therefore, the first step of our study is to validate the model by comparing the networks obtained from one model run with the networks obtained from reanalysis data \citep{deza2013Inferring}.

This can be done by comparing panels (a) and (b) in Fig. \ref{fig:reanalysis} with panels (a) and (d) in Fig. \ref{fig:full}. Figure \ref{fig:reanalysis}(a) displays the AWC map computed from reanalysis data using MIH as interdependency measure; Fig. \ref{fig:full}(a) displays the AWC map computed from one model run, also using MIH. Clearly, the model is able to capture the same overall pattern of global connectivity with a maximum in the central tropical Pacific, relative maxima in the tropical Atlantic and Indian oceans and over Alaska, Labrador Sea and the Southern ocean. Differences are mainly in the magnitude of the AWC, with the model underestimating the connectivity in most places. Similar observation applies to the comparison between Fig. \ref{fig:reanalysis}(b) and Fig. \ref{fig:full}(d) in which the network is built by using the MI OP as interdependency measure, tuned to inter-annual timescales.

In Fig. \ref{fig:full}, panel (a) shows the AWC using MIH and thus, reveals global interdependencies, of all the time-series; panels (b)-(d) show the AWC using MI OP in intra-seasonal, intra-annual and inter-annual time-scale respectively (the ordinal patterns are defined using lags, as explained in Sec. III A). Clearly, the connectivity increases as the time scale increases, in good agreement with the results found in Ref.\citep{deza2013Inferring} using reanalysis data. Many other features of the AWC maps are also qualitatively well reproduced by the model. The main differences with AWC maps from reanalysis are in intraseasonal time scales, in which, while the reanalysis shows a relative maximum of connectivity in the tropical Indo-Pacific region \citep{deza2013Inferring}, the model tends to have a minimum. This indicates that the model has weak tropical intraseasonal variability probably due to the lack of a well represented Madden-Julian Oscillation, due to the coarse resolution and the absence of air-sea interaction \citep{zheng}.

Having observed a good qualitative agreement between the networks constructed from a model run and those constructed from reanalysis data, let us next focus on using the model data to distinguish the influence of forced and internal variability in climate networks.

\subsection{AWC maps}
\subsubsection{Forced variability}

The AWC maps presented in Fig. \ref{fig:full}, for one run of the model, contain information of both forced and internal variability. To analyze \emph{forced variability} only, we have constructed the network from averaged time series (over nine model runs), as explained in the introduction.

The results are presented in Fig. \ref{fig:awcmean}. Panel (a) displays the AWC map when the MIH is used to quantify statistical interdependencies. Here, connectivity is higher in the tropics and on the Pacific, Indian and Atlantic basins. Note that while tropical connectivity is relatively symmetrical about the equator for Pacific and Indian oceans, the north Atlantic is significantly more connected than south of the equator. Panels \ref{fig:awcmean}(b-d) show that the connectivity of the forced variability increases with the time scale. On intra-seasonal time scales connectivity is very low compared with the connectivity from Fig. \ref{fig:awcmean}(a), partly because of the absence of the Madden-Julian Oscilation in the model. If we increase the time scale to intra-annual -- as in panel \ref{fig:awcmean}(c) -- all the tropical area becomes more connected than the extra tropics, indicating a better latitudinal energy and momentum exchange. Forced by the tropical Pacific SST anomalies a long-range strong teleconnection is found in Alaska\citep{Ropelewski1987North}. For inter-annual timescales (three years) which is within the period of the El Ni\~no events (from 2 to 7 years) many very connected areas, especially in the tropics but also in the extratropics are found. The presence of highly connected spots is observed in the extratropics especially in the Pacific basin but also in the Indian and Atlantic oceans. Comparing these three maps with that in panel \ref{fig:awcmean}(a) which, as explained before, was computed via MIH and thus contains information from all the time-series, it can be inferred that most of the connections seen in Fig. \ref{fig:awcmean}(a) occur in long time scales, because they are clear only in Fig. \ref{fig:awcmean}(d), and are weak or not seen in Figs. \ref{fig:awcmean}(b), (c).

Figure \ref{fig:noninomean} represents the same maps as Figure \ref{fig:awcmean} but after removing the NINO3.4 index, as explained in Sec. IV.  Panels \ref{fig:awcmean}(a) and \ref{fig:noninomean}(a) show large differences. It is clear that the signal of El Ni\~no in the tropical Pacific was successfully removed, as only small well-connected areas remain over the equatorial Pacific. Moreover, connection hotspots in the extratropics were also removed indicating that they were mainly forced by El Ni\~no.

The Caribbean and north Atlantic are the largest regions that maintain a similar AWC even after Ni\~no has been removed. Note, however, that the instantaneous regression does not completely remove the ENSO signal if there is a lag in the response. This is so in the tropical north Atlantic\citep{Ropelewski1987North}, where El Ni\~no affects sea surface temperature through heat flux changes that, given the ocean's heat capacity, take a few months to induce an anomaly. Thus, this might be the reason for the still large connectivity observed in the Caribbean in Fig. \ref{fig:noninomean}(a).

Other areas, like over China and central Asia, which are not related to the El Ni\~no phenomenon show the same connectivity in Figs. \ref{fig:awcmean} and \ref{fig:noninomean}. The fact that areas not related to ENSO do not change when removing the index hints that the statistical test used to fix the network density is robust and allows to compare maps with and without the index. This would be not possible by using fixed density networks.

Panel \ref{fig:noninomean}(b) is very similar to panel \ref{fig:awcmean}(b) except on the absence of a connected (dark blue) area on the Pacific ocean, suggesting that the influence of El Ni\~no on these time scales is very low and restricted to the tropical Pacific. In intra-annual time scales, panel \ref{fig:noninomean}(c) shows the disappearance of many links from the corresponding Fig. \ref{fig:awcmean}(c). This suggests that at this time scale, even if El Ni\~no signal is not as strong as on inter-annual scales, it is already connecting far away tropical and extratropical areas as Alaska\citep{Chiang2002Tropical}. Thus, removing El Ni\~no signal affects very heavily the connectivity of the network. For longer time scales -- shown in panel \ref{fig:noninomean}(d) -- the scenario is similar as for \ref{fig:noninomean}(a) with only a remnant of connectivity in the tropical region.

\subsubsection{Internal variability}

Figure \ref{fig:awcmawc} shows AWC maps of internal variability, computed by averaging the nine AWC maps obtained from the individual model runs, where in each time-series, the forced signal (the average of the nine runs) was removed as explained in the Introduction. Contrary to the forced variability case presented before, in this case the most connected areas are on the extratropics. This is consistent with results of previous figures and indicates that in the tropics the ocean forces the largest portion of atmospheric variability. As the tropical atmosphere cannot sustain horizontal gradients generated by SST anomalies, it induces vertical movements of air, convection and release of latent heat, thus giving rise to atmospheric circulation anomalies.

In the extratropics internal atmospheric variability is larger leading to stronger connections. The larger connectivity in the northern hemisphere suggests that the large landmasses affect atmospheric variability, which is consistent with our current understanding of storm track dynamics and low frequency transients.

The most connected spot on Fig. \ref{fig:awcmawc}(a) is over the Labrador Sea. The rest of the highly connected areas (in green) are present mostly in the northern hemisphere. On the southern hemisphere connectivity is largest over the Southern ocean. Investigation over this well connected area --only found using MIH to quantify interdependencies-- showed that in this area histograms have a higher skewness than in the rest of the nodes. This effect is found on the internal-plus-forced AWC map of Fig. \ref{fig:full}(a) and using reanalysis data as shown in panel Fig. \ref{fig:reanalysis}(a). When considering other measures to quantify interdependencies, such as Pearson cross correlation or MI OP, the AWC maps do not show high connectivity in this region \citep{deza2013Inferring}.

With respect to the AWC maps computed by using MI OP, in contrast to the forced case, the intra-season, intra-annual and inter-annual maps are very similar to each other. This is a sign of multiscale variability, distributed over many time scales. Internal variability cycles are less well defined, with spectra similar to ``red" noise. It can be seen that the most connected AWC map is the intra-annual one, stronger than both the intra-seasonal and the inter-annual, consistent with the fact that atmospheric anomalies are less persistent than oceanic ones.

\subsection{Node connectivity maps}

AWC maps provide information of the connectivity of the geographical regions, but no information about the nature --spatial range or distribution-- of the links. It is expected that nearby points behave similarly and this leads to high values of correlation between nearby places. The distance over which the climate variables are well connected is related with the Rossby radius of deformation (RRD)\citep{EncycWorldClim}, which is the distance that a particle or wave travels before being significantly affected by the Earth's rotation. Also, in the tropics, this proximity effect can be greatly enhanced as there the information is propagated very fast longitudinally. Here we are interested in unveiling the presence of teleconnections, that is, connections between regions separated more than the RRD.

The following plots display the connections of a node, indicated with ``X''. Figures display MIH in the left column and MI OP tuned to the time scale involved in the right column. This time scale will be inter-annual for the forced variability network and intra-annual for the internal variability network, as found above. As explained in Sec. III.C, since we are interested in unveiling weak but long-range significant links, we saturated the color scale for nearby links. In this way we are able to see the weak links with good resolution, loosing information for the stronger links (stronger than 0.3) which will be all represented with the same color. 
\subsubsection{Forced variability}

Figure \ref{fig:ninomean} shows the connections of a point in the central Pacific ocean in the forced variability network. The first column shows maps generated from MIH (containing information of all the time series) and the second column, from MI OP inter-annual. It is clear from the comparison of the maps in the first row that most links are inter-annual links.

Panels  \ref{fig:ninomean}(c) and  \ref{fig:ninomean}(d) display the same node connectivity maps, but now the NINO3.4 index has been removed from the time-series and thus (to first order) do not contain links due to El Ni\~no phenomenon. The differences between panels (a) and (c) and between (b) and (d) are evident. First, after eliminating the effects of El Ni\~no the tropical and extratropical teleconnection patterns associated to the spot in the Pacific disappear independently of the methodology used to quantify interdependencies (MIH or MI OP): the connectivity becomes restricted to the tropical Pacific basin. Even inside this region the connectivity is greatly decreased as seen by a much smaller red spot of links over 0.3, although the remaining connections may reflect ENSO dynamics that is not represented by the NINO3.4 index. Notice that the spot is ellipsoidal with a longer longitudinal axis, reflecting the existence of a tropical waveguide. There are still some statistically significant extratropical teleconnections on the map, but are very noisy.

According to Fig. \ref{fig:ninomean} (a) and (b) Alaska is an area well connected to the equatorial Pacific ocean. To further investigate, Fig. \ref{fig:alaskamean} shows global connections to a point nearby Alaska. It can be seen in panels \ref{fig:alaskamean}(a) and \ref{fig:alaskamean}(b)  that it indeed presents connections to the equatorial Pacific ocean with a maximum close to the dateline.

Furthermore, connections to the southern Pacific ocean, Central Africa, Indian ocean and even the Drake passage are found. These connections are stronger on panel (b) especially those linking Alaska with the Indian and southern Atlantic ocean and Drake Passage. If we remove NINO3.4 we find a dramatic change in the maps. Connections become almost local and all the north - south teleconnections are lost; only connections probably associated with an imperfect removal of the El Ni\~no signal remain. This indicates that there are no direct teleconnections between Alaska and (for example) the Drake Passage, but both are strongly connected to El Ni\~no. As these networks are constructed using symmetrical measures of dependency, calculated directly from the data, they are unable to distinguish between a direct connection and an indirect one.

Figure \ref{fig:newZmean} is as Figs. \ref{fig:ninomean} and \ref{fig:alaskamean}, but for a node in the southern hemisphere extratropics. We chose southern New Zealand because it shows a relatively high forced density [seen in Fig. \ref{fig:awcmean} (a,b)] and it is connected to the selected point on the tropical Pacific of Fig. \ref{fig:ninomean} (a,b). Panel \ref{fig:newZmean}(a) shows connectivity between the chosen point and the Pacific and Indian oceans, as long as wave patterns (probably a Rossby wavetrain) along the extra-tropics. Figure \ref{fig:newZmean}(b) adds information to \ref{fig:newZmean}(a) showing that these teleconnections are of inter-annual type. If we remove NINO3.4 (panels \ref{fig:newZmean}(c) and \ref{fig:newZmean}(d)) not surprisingly the links to the tropical Pacific disappear, but also some of the connectivity to the Indian ocean suggesting that part of the links with the Indian ocean are indirect. Nevertheless, the extra-tropical wavetrain remains, and Fig. \ref{fig:newZmean}(d) suggests that the wave train may be forced by the Indian ocean on interannual time scales. As in the previous figure, some weak north-south teleconnections are found, but they disappear if we remove NINO3.4 index, indicating again an indirect connection between the extratropics by means of the Pacific ocean.

\subsubsection{Internal variability}

Figure \ref{fig:labramawc} displays the \emph{internal} variability connections of a node over the most connected area of Fig. \ref{fig:awcmawc}. As with calculating internal variability awc maps, nine different runs are made and  the average of the resulting nine connectivity maps are shown. In the left column we again show the connectivity computed by using MIH while on the right column, by using MI OP tuned at intra-annual scale, as it shows the strongest values of the three time scales considered. In Fig. \ref{fig:labramawc}(a) the original internal variability connections are shown, revealing  teleconnections extending over the northern hemisphere, especially over Scandinavia, Mediterranean Europe, east coast of North America and tropical north Atlantic. Figure \ref{fig:labramawc}(a) also shows connections to eastern China and the Aleutian islands. The pattern shown in Fig. \ref{fig:labramawc}(b) mainly corresponds to the known influence of the North Atlantic Oscillation. This is further substantiated in panels (c) and (d) of the same figure, where the NAO influences is removed and the connections of the Labrador sea, particularly in the northern Atlantic basin, are strongly weakened.

\conclusions[Summary  and conclusions]
We have decomposed the monthly variability of the surface air temperature field into one forced by the ocean temperature and one due to intrinsic atmospheric variability using an ensemble of nine AGCM runs forced with the same SST data starting from slightly different initial conditions. We first validated the model data by observing a good qualitative agreement between the networks constructed from a model run and those constructed from reanalysis data. Then, we constructed climate networks from model data, for the forced and for the internal variability components, using the Mutual Information to assess the interdependencies between time-series. Ordinal patterns have been used in order to separate and determine the strength of the links in different time scales. We also generated time series where the evolution of well known phenomena (El Ni\~no and NAO) have been removed.

We found that forced and internal atmospheric variability are characterized by very different networks and thus possess different properties which cannot be inferred from observational/reanalysis data. The network of forced variability has the strongest connections in interannual time scales on the tropics with long range teleconnections to the extratropics mainly due to the influence of El Ni\~no. The presence of \emph{long-range teleconnections} between different hemispheres in the forced network are indirect and due to the influence of El Ni\~no. On the other hand, the network of internal atmospheric variability has strongest connections in the extratropics, and the North Atlantic Oscillation is the most important connector globally.

The methodology proposed here for distinguishing links in spatial range (short and long), time scale (intra-season, intra-annual and inter-annual) and type of variability (forced vs. internal) is a novel approach for the study climate networks that provides new insight in the climatological meaning of the links found and their connection to physical phenomena.

\begin{acknowledgements}
The research leading to these results has received funding from the European
Community's Seventh Framework Programme FP7/2007-2011 under grant agreement n$\circ$ 289447.
\end{acknowledgements}

\end{document}